\documentclass[twocolumn]{aastex631}

\shorttitle{The Wanderer: Charting Hot Jupiter Formation and Migration}
\shortauthors{Coria et al.}

\begin{document}

\title{The Wanderer: Charting WASP-77A b's Formation and Migration Using a System-Wide Inventory of Carbon and Oxygen Abundances}

\author[0000-0002-1221-5346]{David R. Coria}
\affiliation{Department of Physics $\&$ Astronomy, University of Kansas, Lawrence, KS, USA}
\email{Email: drcoria@ku.edu}

\author[0000-0001-5541-6087]{Neda Hejazi}
\affiliation{Department of Physics $\&$ Astronomy, University of Kansas, Lawrence, KS, USA}

\author{Ian J. M. Crossfield}
\affiliation{Department of Physics $\&$ Astronomy, University of Kansas, Lawrence, KS, USA}

\author{Maleah Rhem}
\affiliation{Department of Physics $\&$ Astronomy, University of Kansas, Lawrence, KS, USA}



\begin{abstract}

The elemental and isotopic abundances of volatiles like carbon, oxygen, and nitrogen may trace a planet’s formation location relative to H$_2$O, CO$_2$, CO, NH$_3$, and N$_2$ “snowlines”, or the distance from the star at which these volatile elements sublimate. By comparing the C/O and $^{12}$C/$^{13}$C ratios measured in giant exoplanet atmospheres to complementary measurements of their host stars, we can determine whether the planet inherited stellar abundances from formation inside the volatile snowlines, or non-stellar C/O and $^{13}$C enrichment characteristic of formation beyond the snowlines. To date, there are still only a handful of exoplanet systems where we can make a direct comparison of elemental and isotopic CNO abundances between an exoplanet and its host star. Here, we present a $^{12}$C/$^{13}$C abundance analysis for host star WASP-77A (whose hot Jupiter's $^{12}$C/$^{13}$C abundance was recently measured). We use MARCS stellar atmosphere models and the radiative transfer code TurboSpectrum to generate synthetic stellar spectra for isotopic abundance calculations. We find a $^{12}$C/$^{13}$C ratio of $51\pm 6$ for WASP-77A, which is sub-solar ($\sim 91$) but may still indicate $^{13}$C-enrichment in its companion planet WASP-77A b ($^{12}$C/$^{13}$C = 26 $\pm$ 16, previously reported). Together with the inventory of carbon and oxygen abundances in both the host and companion planet, these chemical constraints point to WASP-77A b's formation beyond the H$_2$O and CO$_2$ snowlines and provide chemical evidence for the planet’s migration to its current location $\sim$0.024 AU from its host star.  

\end{abstract}

\keywords{Planet hosting stars --- Exoplanet Systems --- Hot Jupiters --- Exoplanet Migration --- Exoplanet formation --- Exoplanet atmospheric composition --- Elemental abundances --- Isotopic abundances}


\section{Background} \label{sec:bg}
\subsection{Refractory Abundances and the Star-Planet Connection}
Stellar composition has long been implicated in planet formation-- after all, stars and planets form out of the same disk of gas and dust albeit through different mechanisms. Different protoplanetary disk compositions and planet formation mechanisms have diverse outcomes which produce rocky terrestrial planets, super-Earths, sub-Neptunes, and even gas giants like Jupiter and Saturn. Since stellar atmospheres evolve slowly, the elemental abundances of exoplanet hosts tend to reflect the composition of their planet-forming disks \citep{Brewer_2016} and have the potential to yield constraints on planet formation processes and, in turn, even the physical properties of exoplanets themselves \citep{Bedell_2018}. One of the best known links between stellar abundances and exoplanets is stellar metallicity. The occurrence rate of close-in ($<$1 au) Jupiter-class planets is increased around higher metallicity stars \citep{Valenti_Fischer_2005, Johnson_2010, Mortier_2013}. This relation is interpreted as support for the core accretion model: protoplanetary disks around more metal-rich stars often have higher masses and metal contents  \citep{Andrews_2013} and can sustain solids for longer periods \citep{Bitsch_Mah_2023}, which allows giant planets to form more efficiently. Planet formation is therefore  predicted to be unlikely in in metal-poor environments lacking sufficient metals to form planetary cores and kick-start accretion \citep{Andrews_2013, Boley_2021, Bitsch_Mah_2023}. The relation with host-star metallicity, though weaker, has also been observed for sub-Neptunes and Super-Earths \citep{Wang_Fischer_2015, Petigura_2018, Boley_2024}. Besides stellar metallicity, there is also a correlation between the abundances of individual refractory elements like (Mg, Si, Al, Ti) and planetary occurrence rates \citep{Adibekyan_2012}. For example, giant-planet host stars exhibit higher refractory-to-iron abundance ratios than non-host stars. In Neptune-class planet hosts, there is notable increase in [Ti/Fe], [Si/Fe], [Al/Fe], and [Mg/Fe] compared to non-host stars. 

Refractory-to-volatile ratios also become useful in tracing giant exoplanet formation and migration for the hot and ultra-hot Jupiter-class exoplanets \citep{Hands_Helled_2022, Lothringer_2021}. In these planets with equilibrium temperatures greater than 2000 K, refractory elements like Fe, Mg, and Si are not condensed, except perhaps on the night side, but rather gaseous and observable in the planet's atmosphere \citep{Lothringer_2021}. Using these atmospheric refractory-to-volatile abundance ratios, we may infer a planet's rock-to-ice fraction and constrain planet formation and migration scenarios \citep{Lothringer_2021}. Furthermore, formation of a planet beyond the water snowline followed by inward migration results in excess accretion of oxygen-poor, refractory-rich material leading to super-stellar alkali metal abundances but stellar water abundances in the planet's atmosphere \citep{Hands_Helled_2022}.

\subsection{Volatile Element Abundances as Formation Diagnostics}\label{elemental_ab_diagnostics}
Similarly, the stellar abundances of volatile elements like H, C, N, O, and S may shed light on giant exoplanet formation mechanisms, planet composition and atmospheric evolution \citep{Bedell_2018, Brewer_Fischer_2016, Fortney_2012, Delgado_Mena_CO_2021, Turrini_2021, Crossfield_2023}. Stellar carbon and oxygen abundances are particularly important formation diagnostics because of their influence on exoplanet ice, gas, and rock chemistry. Formation mechanisms such as gravitational instability for brown dwarfs and core accretion for Jupiter-class exoplanets each leave behind unique carbon and oxygen signatures in the atmospheres of these sub-stellar objects. Formation via gravitational instability occurs more rapidly than core accretion, on sub-Myr timescales, which results in brown dwarfs with stellar to super-stellar C/O ratios and abundances characteristic of their progenitor molecular clouds, similar to binary star systems \citep{Ilee_2017, Hawkins_2020}. Core accretion, on the other hand, is a much slower process, occurring on Myr timescales, which allows for protoplanets to incorporate varying quantities of gas and solids into their atmospheres, potentially resulting in a wide range of atmospheric metallicities and C/O ratios, but maintaining sub-stellar carbon and oxygen abundances \citep{Xuan_2024b}.

Multiple studies also explore the possibility of using elemental abundance ratios like the carbon-to-oxygen ratio \citep{Oberg_2011, Madhusudhan_2014, Mordasini_2016, Ali-Dib_2017, Reggiani_2022, Seligman_2022} to constrain a planet's formation location relative to ``snowlines", or the distance from a star where volatile molecules like H$_2$O, CH$_4$, CO, and CO$_2$ condense into ice grains. When a planet forms within the H$_2$O and CO snowlines, like Jupiter and Saturn for example, the planet is expected to inherit the same C/O ratio as its host star. When the planet forms outside the volatile snowlines, there are two main scenarios that affect the volatile abundances of these planets: (1) the planet migrates before dissipation of the protoplanetary disk resulting in atmospheric C/O ratios $< 0.5$ and super-solar metallicities due to the accretion of oxygen-rich planetesimals or (2) the planet migrates after dissipation of the protoplanetary disk resulting in atmospheric C/O ratios $\sim 1$ and sub-solar metallicities \citep{Oberg_2011, Madhusudhan_2014, Madhusudhan_2017, Madhusudhan_2019}. See also \citep{Schneider_Bitsch_2021a, Schneider_Bitsch_2021b, Bitsch_2022, Pacetti_2022} for an in-depth analysis on how inward drifting evaporating pebbles affect volatile abundances in exoplanet atmospheres along with a couple example giant exoplanet formation scenarios. 

Though precise stellar abundances provide useful context for interpreting the atmospheric composition of their companion exoplanets \citep{Reggiani_2022, Reggiani_2024}, there is still a need to investigate the connection between atmospheric composition, protoplanetary disk chemistry, and planetary formation inferences \citep{Molliere_2022}. After all, the chemical composition of giant planets and their atmospheres are greatly affected by the accretion of gas and solids as they form and migrate to their final locations, deviating significantly from abundances of the protoplanetary disk \citep{Pacetti_2022}. Comparisons of the atmospheric abundances of exoplanets to planet formation simulations provide additional, complementary constraints on planet formation pathways. The condensation sequence of H$_2$O, CO$_2$, CH$_4$, and CO results in an increase of the gas phase C/O ratio with radial distance from the host star, which makes the planetary C/O ratio an ideal signature of a planet's accretion history \citep{Bitsch_2022}. Besides C/O, other volatile ratios like C/N, N/O, and S/N appear to follow monotonic trends with the extent of migration-- the deviation from stellar values increases with the extent of disk-driven migration \citep{Pacetti_2022}. By measuring these volatile element abundances in an exoplanet's atmosphere, comparing them to the host star's abundances as well as planet formation models, we may predict how and where in the disk the planet formed and infer subsequent migration. 

\subsection{Isotopologue Abundance Ratios as Formation Diagnostics}\label{isotopic_ab_diagnostics}
In the era of the James Webb Space Telescope (JWST), ground-based 8m- class telescopes, upcoming 30m- class telescopes, and ultra high-resolution spectroscopy, we are suddenly sensitive to the signatures of much weaker molecular absorption lines in dwarf stars, brown dwarfs and even exoplanets. In addition to the major isotopologues of CO, CH, H$_2$O, NH$_3$, we are now detecting minor isotopologues such as $^{13}$C$^{16}$O, $^{12}$C$^{18}$O, $^{13}$CH, H$_2$$^{17}$O, H$_2$$^{18}$O, and $^{15}$NH$_3$ which allow for the measurements of $^{12}$C/$^{13}$C, $^{14}$N/$^{15}$N, $^{16}$O/$^{18}$O, and $^{17}$O/$^{18}$O isotopologue ratios. The most widely observed of the CNO isotope ratios is the $^{12}$C/$^{13}$C ratio. Measurements have been made in young stellar objects \citep{Smith_2015}, solar twin stars \citep{Botelho_2020, Coria_2023}, M dwarf stars \citep{Crossfield_2019, Xuan_2024}, brown dwarfs (e.g. L dwarf 2M0355, T dwarf, 2M0415, M dwarf HD 984 B, and several others) \citep{Zhang_2021b, Costes_2024, Hood_2024, Xuan_2024b, de_Regt_2024}, short-orbit hot-Jupiters (e.g. WASP-77A b and HD 189733 b) \citep{Line_2021, Finnerty_2024} and wide-orbit Jupiters (e.g. TYC-8998 b and VHS 1256 b) \citep{Zhang_2021a, Gandhi_2023} using instruments and observatories such as VLT/CRIRES, ESO/HARPS, NASA IRTF/iSHELL, Keck/KPIC, Keck/NIRSpec, Gemini/IGRINS, and JWST/NIRSpec. 

These isotopic ratios are not only useful in constraining Galactic Chemical Evolution (GCE) models \citep{Kobayashi_2020, Prantzos_2018, Romano_2020, Romano_2022, Coria_2023}, but also in understanding planet formation pathways and accretion histories \citep{Woods_Willacy_2009, Atreya_2016, Zhang_2021a, Zhang_2021b, Barrado_2023}. The first study investigating carbon isotope fractionation in protoplanetary disks finds that the $^{12}$C/$^{13}$C ratio of a system varies with radius and height in the disk \citep{Woods_Willacy_2009}. Complementary observations and measurements of the $^{12}$C/$^{13}$C ratio in solar system objects provide a hint of $^{13}$C-enrichment of the ice giants located beyond our Sun's CO snowline \citep{Woods_2009}. Since then, observations of $^{12}$C/$^{13}$C in protoplanetary disks have unveiled different isotopic reservoirs that may imprint their signatures in exoplanet atmospheres \citep{Zhang_2017, Nomura_2023, Bergin_2024, Yoshida_2024}. Beyond the CO snowline at around 20 AU for sun-like stars, \cite{Bergin_2024} finds one reservoir enriched in $^{12}$C composed mainly of methane/hydrocarbon ices and a second, $^{13}$C-enriched reservoir dominated by gaseous CO. \cite{Zhang_2021a} instead proposes $^{13}$C-enriched ice, found beyond the CO snowline, as a source for observed giant planet $^{13}$C enrichment. 

Similar to elemental carbon and oxygen abundances, different formation mechanisms also produce signature atmospheric $^{12}$C/$^{13}$C ratios. The sub-Myr timescales of formation via gravitational instability result in brown dwarfs with solar to super-solar $^{12}$C/$^{13}$C ratios while the longer, Myr timescales of formation via core accretion allow for protoplanets to accrete $^{13}$C-rich ice/gas. Although there are few planetary growth models that incorporate isotopolgue ratios, this particular pattern is arising from $^{12}$C/$^{13}$C ratio measurements made in brown dwarfs (e.g. L dwarf 2M0355, T dwarf, 2M0415, M dwarf HD 984 B) \citep{Zhang_2021b, Costes_2024, Hood_2024}, and lower-mass super-Jupiters (e.g. WASP-77A b, HD 189733 b, TYC-8998 b and VHS 1256 b) \citep{Line_2021, Finnerty_2024, Zhang_2021a, Gandhi_2023}. Furthermore, the ice-gas partitioning of these carbon isotopes allows for the $^{12}$C/$^{13}$C ratio to be used as a tracer of a planet's formation location and migration. Planets forming within the CO snowline are expected to inherit the host star's $^{12}$C/$^{13}$C ratio while those forming beyond the CO snowline (~20 AU) are expected to accrete $^{13}$C-rich ice/gas which lower the planet's $^{12}$C/$^{13}$C ratio. The formation scenarios and corresponding C/O and $^{12}$C/$^{13}$C ratios outlined above assume in-situ formation. However, if there is a mismatch between a planet's C/O ratio, $^{12}$C/$^{13}$C ratio, its current location, and the host star's C/O and $^{12}$C/$^{13}$C ratio, this could be an indicator for planetary migration. Recent observations of both hot Jupiters (e.g. WASP-77A b and HD 189733 b) \citep{Line_2021, Finnerty_2024} and wide-orbit, directly imaged planets (e.g. VHS 1256 b and TYC 8998-760-1 b) \citep{Gandhi_2023, Zhang_2021a} reveal similar $^{13}$C enrichment, supporting the hypothesis that hot Jupiters likely did not form in situ.

To test how planetary $^{12}$C/$^{13}$C ratios change with distance with their host stars and relative to various volatile snowlines, it is necessary to measure volatile abundances in three types of targets: (1) planets located inside the CO snowline, (2) planets located outside the CO snowline, and (3) exoplanet host stars. Jupiter-class planets are the most amenable to these system-wide abundance surveys. Short-period super-Jupiters close to their host stars (within the CO snowline) can have their C/O and $^{12}$C/$^{13}$C ratios measured using transmission spectroscopy (as in \cite{Line_2021}) while bright super-Jupiters outside the CO snowline, like TYC-8998-760-1 b \citep{Zhang_2021a}, can have their C/O and $^{12}$C/$^{13}$C ratios measured using a combination of spectroscopy and direct imaging techniques \citep{Molliere_2019}. Although FGK-type host stars routinely have their abundances measured using standard methods \citep{Brewer_Fischer_2016, Bedell_2018}, cooler mid- to late- type K and M dwarfs pose a bigger challenge due to their intrinsic faintness and overwhelming molecular absorption in their spectra. However, with high-resolution, high signal-to-noise spectra and a careful analysis we may successfully measure abundances in these stars as well \citep{Souto_2017, Souto_2018, Souto_2022, Hejazi_2023, Hejazi_2024}. Contemporary researchers use both the stellar and planetary abundances to model exoplanet atmospheres, to infer the structure and composition of the exoplanet's interior, and to understand how atmospheres and interiors co-evolve over time \citep{Madhusudhan_2012, Unterborn_2014, Brewer_2016, Unterborn_2017, Lincowski_2018, Lincowski_2019}. 

In this paper, we focus on the complementary stellar and planetary isotopic abundances of the volatile elements: carbon and oxygen. We present the $^{12}$C/$^{13}$C ratio measured in exoplanet host star WASP-77A using optical CH absorption features. We then compare archival carbon and oxygen abundances and our novel $^{12}$C/$^{13}$C ratio in WASP-77A to those in its companion, the hot-Jupiter WASP-77A b. Our isotope ratio measurement corroborates previous findings from stellar and planetary elemental abundance analyses that favor formation beyond the H$_2$O and CO$_2$ snowlines, rather than an in-situ formation, for WASP-77A b.

In Section \ref{sec:obs}, we provide an overview of three previous elemental abundance analyses for the host star (WASP-77A) with data from Keck/HIRES, MPG/ESO/FEROS, and ARC/ARCES. We also summarize the constraints placed on the companion planet's (WASP-77A b) C/O and $^{12}$C/$^{13}$C abundance ratios derived from thermal emission spectroscopy. WASP-77A b is a well-studied hot Jupiter, so we include the companion's abundance measurements using data from three facilities: Gemini/IGRINS, HST/WFC3, and JWST/NIRSpec. 

Section \ref{sec:methods} provides details on our Keck/HIRES and VLT/ESPRESSO spectra, and our methodology for determining $^{12}$C/$^{13}$C for the host star WASP-77A. In Section \ref{sec:discussion}, we discuss WASP-77A b's formation scenario and the possible migration mechanisms that may have transformed a once cool-Jupiter to the hot-Jupiter we see today. Finally, in Section \ref{sec:conclusions}, we summarize the results of our stellar $^{12}$C/$^{13}$C analysis and discuss the prospects for measuring more complementary stellar isotopologue ratios.

\section{Previous Observations of the WASP-77 System} \label{sec:obs}
The WASP-77 system is a visual binary otherwise known as BD -07 436. The primary star (WASP-77 A) is a moderately bright, solar metallicity G8V star (V$_{mag} = 10.3$) and the secondary star (WASP-77 B) is a fainter K dwarf at a separation of approximately 3$\arcsec$. \cite{Maxted_2013} discovered a transiting, tidally-locked planetary mass companion (WASP-77A b) to the primary with an orbital period of 1.36 days. WASP-77A b is slightly larger than Jupiter with a mass of $1.76 \pm 0.06\ M_{Jup}$ and radius of $1.21 \pm 0.02\ R_{Jup}$ and rather hot with an effective temperature of 1740 K\citep{Maxted_2013}. Both WASP-77 A and its companion planet are well-studied, well-characterized, and prime targets for chemical composition analyses as summarized below. 

\subsection{Host Star: WASP-77A High-resolution Spectroscopy with Keck/HIRES, MPG/ESO/FEROS, \& ARC/ARCES}
\cite{Polanski_2022} uses high-resolution, high signal-to-noise optical spectra from Keck/HIRES and KeckSpec and a machine-learning-based tool to derive fundamental stellar parameters and abundances for 25 prime JWST target exoplanet host stars. Please refer to Table \ref{tab:abundances} for their adopted effective temperature, surface gravity, and [Fe/H]. Although they provide stellar abundances for 15 elements, here we include only the constraints placed on WASP-77A's carbon, oxygen, and [Fe/H] abundances. They report stellar [C/H] = $-0.02 \pm 0.05$, [O/H] = $0.06 \pm 0.07$, C/O = $0.46 \pm 0.09$, and [Fe/H] = $+0.01 \pm 0.03$.

\cite{Kolecki_2022} works with an MPG/ESO/FEROS spectrum (S/N = 116) to derive chemical abundances of elements most informative to planet formation and composition (C, N, O, Na, Mg, Si, S, K, and Fe). Here, we include only the stellar [C/H] = $-0.04 \pm 0.04$, [O/H] = $-0.04 \pm 0.04$, C/O = $0.59 \pm 0.08$, and [Fe/H] = $-0.15 \pm 0.06$. Again please refer to Table \ref{tab:abundances} for the adopted stellar parameters. Notice that although they measure carbon and oxygen abundances consistent with solar values, they do, however, find a slightly subsolar [Fe/H].

\cite{Reggiani_2022} also determined the effective temperature, surface gravity, and [Fe/H] for WASP-77A using an optical spectrum (from ARC/ARCES at the Apache Point Observatory), ATLAS9 model atmospheres \citep{Castelli_Kurucz_2003}, the MOOG radiative transfer code \citep{Sneden_1973, Sneden_2012}, and the \textit{isochrones} package \citep{Morton_2015}. Please refer to Table \ref{tab:abundances} for their adopted stellar parameters. They infer elemental abundances for WASP-77A using the equivalent width method and find [C/H] = $0.1 \pm 0.09$, [O/H] = $0.23 \pm 0.02$, and C/O = $0.44 \pm 0.07$. These abundances are mostly consistent with previous publications: \cite{Maxted_2013}, \cite{Kolecki_2022}, and \cite{Polanski_2022}. These abundances for WASP-77A b and its host star are indicative of formation beyond the H$_2$O snowline, rather than in situ. We discuss this further in Section \ref{sec:discussion}.

\subsection{Differences Between Abundance Measurements}
The relevant stellar parameters and abundances discussed above are shown in Table \ref{tab:abundances}. All three studies above find a slightly subsolar iron abundance for WASP-77A and a solar to slightly subsolar C/O ratio. Though mostly consistent, there are a few notable discrepancies in effective temperature, [Fe/H], carbon and oxygen abundance values. Oxygen abundances are notoriously difficult to determine, particularly when using optical spectra where there is a shortage of detectable atomic oxygen lines with both observational and theoretical complications \citep{Ting_2018}. In this case, however, the differences are likely due to the slightly different T$_{eff}$ and [Fe/H] values adopted.

Generally, significant differences in abundance measurements can be caused by a number of factors including the choice of input stellar parameters (T$_{eff}$, log g, [Fe/H]), the selected stellar model atmospheres, the radiative transfer code used for generating synthetic spectra, and the preferred atomic and molecular line lists. We refer the reader to \cite{Kolecki_2022} Section 5.2 and the references therein for a more detailed discussion of abundance discrepancies. For future isotopologue analyses, this means that an elemental abundance fit prior to measuring isotopic abundances is necessary to ensure the best possible fit between the base model spectrum and the observed spectrum as this is the starting point for generating the $^{12}$C/$^{13}$C model grid.

\subsection{WASP-77A b Thermal Emission Spectroscopy}
WASP-77A b is an excellent candidate for transit spectroscopy and is actually one of the highest signal-to-noise ratio planets for thermal emission measurements in the near-infrared \citep{Kempton_2018}-- making this hot Jupiter an ideal target for atmospheric characterization. 

\cite{Line_2021} provides secondary eclipse spectroscopy of WASP-77A b. The single 4.7 hour time-series sequence was taken with Gemini-South/IGRINS \citep{Park_2014, Mace_2018} on December 14, 2020. The resulting spectrum is high-resolution (R $\sim$ 45,000) with wavelength coverage spanning 1.43 - 2.42 $\mu$m.

The IGRINS analysis reports a planetary [C/H] = $-0.46^{+0.17}_{-0.16}$, [O/H] = $-0.49^{+0.14}_{-0.12}$, and a carbon-to-oxygen ratio C/O = $0.59 \pm 0.08$. More importantly, they also retrieve a constraint on the $^{12}$CO/$^{13}$CO ratio: 10.2-42.6 at 68\% confidence. The sub-solar carbon and oxygen abundances in WASP-77A b suggest an atmosphere depleted in metals, at least based on extrapolation from the solar system planets. Although a comparison to solar abundances is a natural first step, the key is to make the comparison with the planet's host star abundances. Refer to Section \ref{sec:discussion} for an abundance comparison between the host star WASP-77A and the companion planet WASP-77A b and see Table \ref{tab:abundances} for a list of all available carbon and oxygen abundances and metallicities for the WASP-77A system.

\cite{Mansfield_2022} presents another set of secondary eclipse observations for WASP-77A b, this time with the Hubble Space Telescope's WFC3 covering wavelengths 1.1-1.7 $\mu$m and Spitzer/IRAC at 3.6 and 4.5 $\mu$m. This atmospheric retrieval places a 3$\sigma$ lower limit on the atmospheric H$_2$O abundance: log(n$_{H2O}$) $> -4.78$ but was unable to constrain the CO abundance and individual [C/H] and [O/H] abundances. There are no carbon-bearing molecule features resolved in this emission spectrum, which results in a poor constraint on the C/O ratio: a 2$\sigma$ upper limit of C/O = 0.78. The retrieval's best fit metallicity is [M/H] = 0.43$^{+0.36}_{-0.28}$. This value is much higher and less precise than that derived from the high-resolution Gemini/IGRINS observations. After performing a grid fit to a combination of the WFC3 and Spitzer data, they derive a metallicity of [M/H] = 0.10$^{+0.43}_{-0.31}$, which is more consistent with the high-resolution measurement. 

Recent secondary eclipse observations of WASP-77A b with JWST/NIRSpec \citep{August_2023} cover the wavelength range 2.8-5.2 $\mu$m and contains several H$_2$O and CO features but no CO$_2$. The atmospheric retrieval finds a sub-solar metallicity [M/H] = $-0.91^{+0.24}_{-0.16}$ and a C/O ratio = $0.36^{+0.10}_{-0.09}$ as well as molecular abundances of log$_{10}$(n$_{H_2O}$) = $-4.26^{+0.14}_{-0.10}$ for water and log$_{10}$(n$_{CO}$) = $-4.58^{+0.27}_{-0.24}$ for carbon monoxide. These results agree with the Gemini/IGRINS abundances from \cite{Line_2021} within $\sim 1\sigma$ for [Fe/H] and $\sim 1.8\sigma$ for the C/O ratio.

\section{Methods: Isotope Analysis} \label{sec:methods}
\subsection{The Spectrum}
We observed WASP-77A with the HIRES spectrograph \citep{Vogt_1994} at the Keck Observatory. The 582-second exposure of WASP-77A was taken on February 22, 2021 06:03 UT by the California Planet Search collaboration (Lead: Andrew Howard, Caltech). They use the C2 decker under effective seeing conditions of roughly 1.3$\arcsec$. Standard reduction routines and telluric corrections are applied resulting in a spectrum of R $\sim$ 45,000 and median signal-to-noise of 82. The final spectrum covers a wavelength range of 3600-4400 $\AA$. This is the same spectrum analysed in \cite{Polanski_2022}. Additionally, we also utilize publicly available ESO data for WASP-77A. This 14400-second exposure of WASP-77A was taken on October 29, 2021 02:03 UTC using VLT/ESPRESSO's ultra-high-resolution (UHR) mode. The data is reduced using the ESPRESSO data-reduction software (DRS) pipeline, and the final extracted, wavelength-calibrated spectrum has R $\sim 140,000$, S/N = 511 and covers a wavelength range of 3770-7898 $\AA$.

\subsection{Generating the Synthetic Spectra Grid}\label{subsec: gen_mod}
We will measure CH isotopologue abundances in WASP-77A by comparing the observed spectrum to synthetic spectra generated from custom 1D hydrostatic MARCS stellar atmosphere models \citep{Gustafsson_2008} and the LTE radiative transfer code TurboSpectrum (Version 15.1) \citep{Plez_2012}. Although there is an extensive grid of MARCS models, we further used the interpolation routine developed by Thomas Masseron \footnote{https://marcs.astro.uu.se/software.php} to interpolate models with the same physical parameters as those of our target star. We also use the set of atomic and molecular line lists, assuming solar abundances from \citep{Grevesse_2007}, as described in \cite{Hejazi_2023}. This includes atomic line data from the Vienna Atomic Line Database (VALD, \cite{Ryabchikova_2015}) and molecular line lists from multiple sources including VALD for TiO, the Kurucz (Smithsonian) Atomic and Molecular Database \citep{Kurucz_1995}, and the high resolution transmission molecular absorption database (HITRAN, \cite{Rothman_2021}). The most important molecular bands for this study are: FeH \citep{Dulick_2003}, CN \citep{Brooke_2014, Sneden_2014}, and CH \citep{Masseron_2014}.

Initially, we adopt the effective temperature, surface gravity, and [Fe/H] (as a proxy for the bulk stellar metallicity) values for WASP-77 A derived from our Keck/HIRES spectrum using the machine-learning tool KeckSpec: T$_{eff} = 5569 \pm 77$ K, log g $= 4.45 \pm 0.09$, and [Fe/H] $= 0.01 \pm 0.03$ \citep{Polanski_2022}. At this point, we have not altered individual elemental abundances from solar values. However, the base model generated from these parameters did not appear to fit the spectral lines surrounding several target $^{13}$CH lines in the HIRES spectrum, particularly for atomic iron lines and molecular CH lines. The problem persists for a couple lines in the higher-quality ESPRESSO spectrum, but the overall fit is greatly improved. This leads us to believe that the issue with these specific lines lies in the opacity calculations or the atomic and molecular line lists incorporated in synthesizing spectra rather than the base stellar parameters. Rather than choose a set of effective temperature, surface gravity, and [Fe/H] (as a proxy for the bulk stellar metallicity) from one particular study, we instead adopt weighted averages (weights = $1/\sigma^2$ where $\sigma$ is the parameter uncertainty) for the stellar parameters since the effective temperature and surface gravity values are similar for the three previous host star studies \citep{Polanski_2022, Reggiani_2022, Kolecki_2022}: T$_{eff} = 5545 \pm 22$ K, log g $= 4.45 \pm 0.02$. Please refer to Section \ref{subsec: sys_uncert} for further explanation of how our choice of fundamental stellar parameters affects the derived $^{12}$C/$^{13}$C ratio.

From these chosen stellar parameters, we then generate a grid of synthetic spectra with $^{12}$C/$^{13}$C ratios of 30, 50, 60 80, 90, 100, 110, 130, 160. These model spectra cover wavelengths 3600 - 4400 \AA\ where there are numerous $^{12}$CH and $^{13}$CH lines suitable for this isotopologue abundance analysis.

\subsection{Line Selection}
We utilize TurboSpectrum's built-in equivalent-width calculation feature to generate a list of equivalent widths of the $^{12}$CH and $^{13}$CH lines from the base model. From that list, we remove all $^{13}$CH lines with EW $< 1.5$ \AA\ and are left with ~95 candidate absorption lines. We narrow down this line list via a visual inspection of our base model spectrum. Lines from the final set are later accepted or rejected from the final $^{12}$C/$^{13}$C calculation depending on overall model fit and continuum renormalization. Please refer to Table \ref{tab:line_list} for our final $^{13}$CH line list. We also include information for atomic and molecular lines located within the renormalization window of our $^{13}$CH lines of interest.

\subsection{Continuum Re-normalization}
After applying a Gaussian filter to smooth the model spectra to a spectral resolution $\sim$45,000 for the HIRES spectrum and $\sim$140,000  for the ESPRESSO spectrum, and performing an RV correction to the observed spectrum, we begin the continuum selection and re-normalization following the methods described in \cite{Hejazi_2023}. The $^{13}$CH features in the optical are significantly weaker than the $^{13}$CO features identified in the M-band (4.5-5.0 \micron) and so this isotopic abundance analysis demands a more precise continuum-selection and re-normalization process than in \cite{Coria_2023}. Otherwise, even a slight misalignment between the observed spectrum and the model grid (either wavelength or continuum normalization misalignment) makes determining the best fit model nearly impossible. 

\begin{figure*}
\epsscale{1.0}
\plottwo{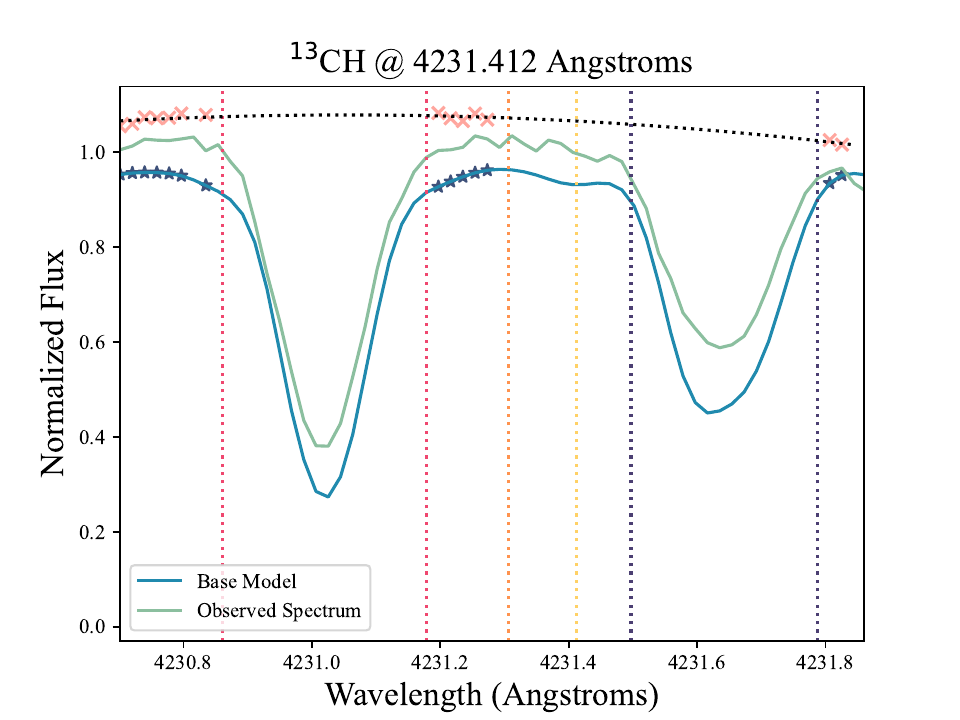}{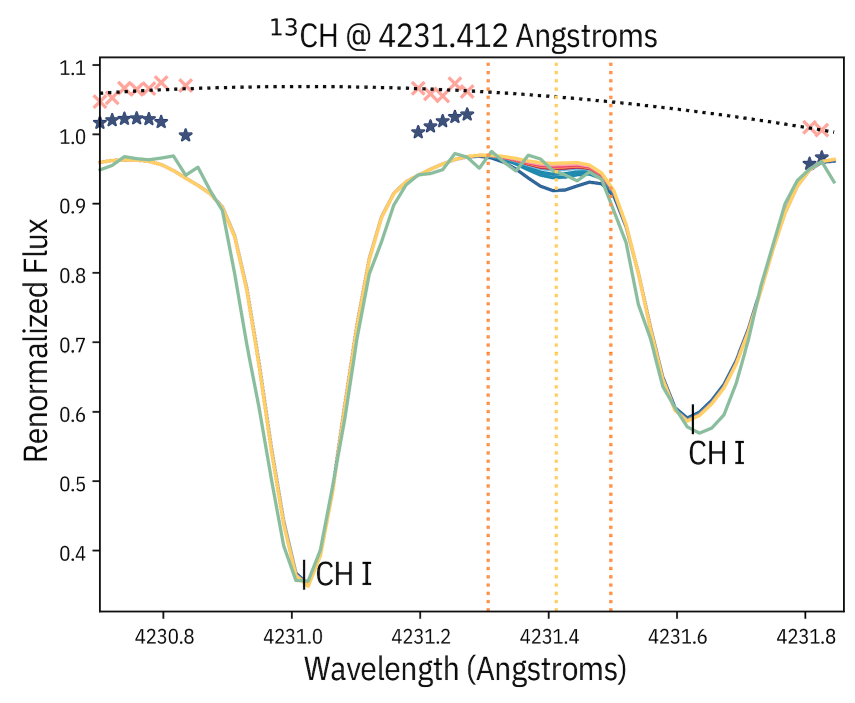}
\plottwo{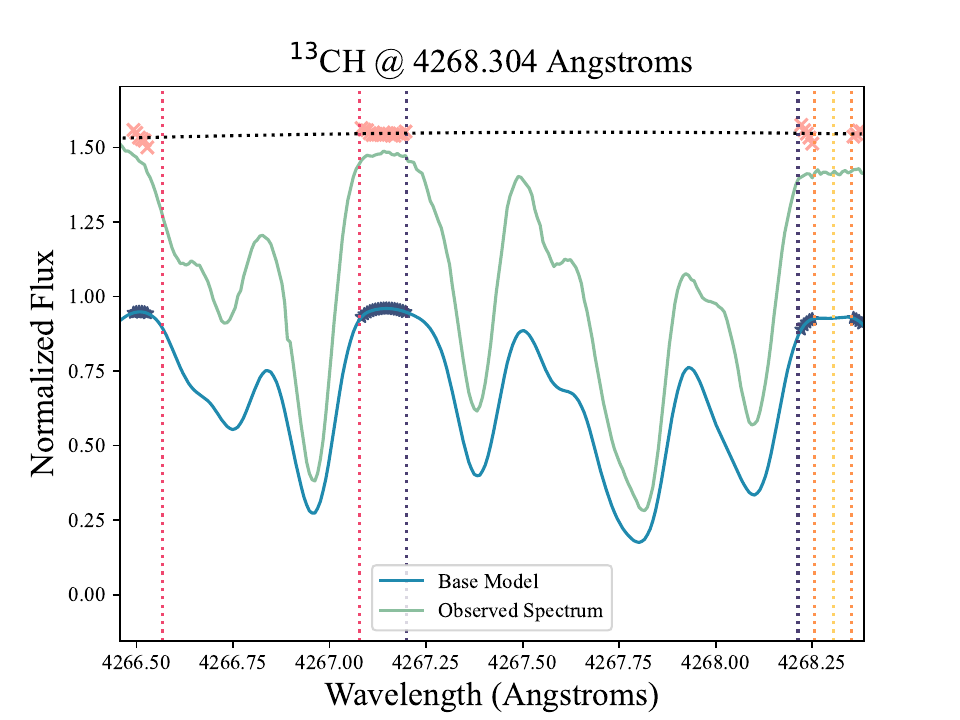}{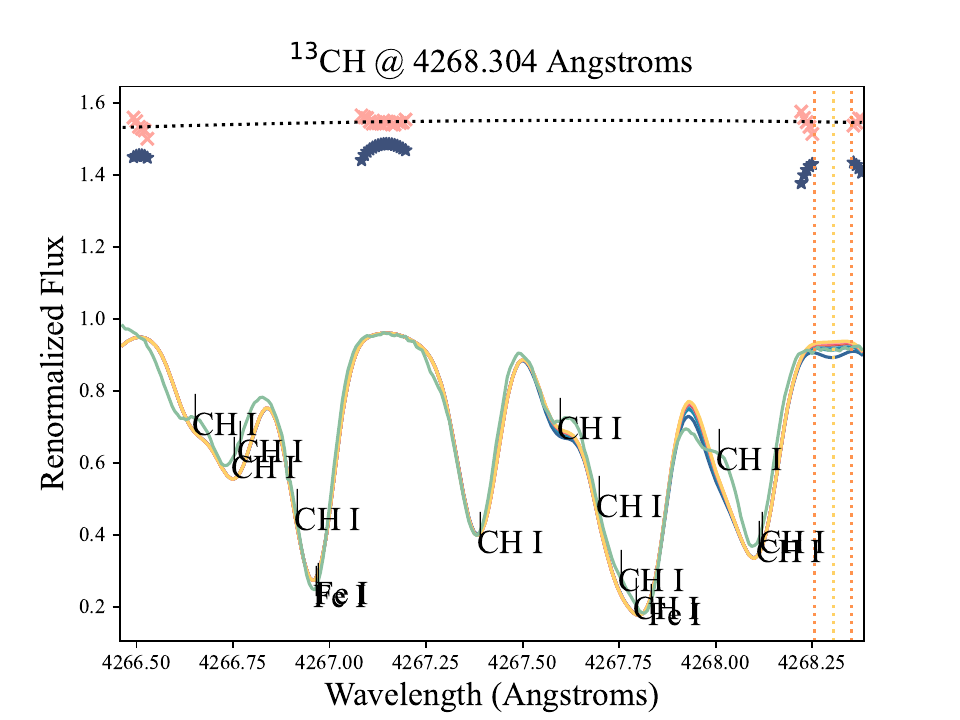}
\caption{Pre- and post- renormalization for strong, isolated $^{13}$CH lines (Top: Line at 4231.412 \AA\ in the HIRES spectrum; Bottom: Line at 4268.304 \AA\ in the ESPRESSO spectrum). The area within the orange lines shows the $\chi^2$ fit window dominated by $^{13}$CH line centered at 3982.247 Angstroms, while the regions covered by the red and black lines (left plot) are excluded from the continuum/pseudocontinuum selection routine. The starred points are the selected continuum/pseudocontinuum points used for renormalization. The pink points represent the residuals R = O/M for the selected continuum/pseudocontinuum points and the black curve is the polynomial fit which, when divided out of the observed spectrum, provides a significantly better alignment to the target $^{13}$CH line. The figures on the right highlight the location of our selected $^{13}$CH lines and other nearby atomic lines.}
\label{pre_re_norm}
\end{figure*}

For each of the selected $^{13}$CH lines, we visually inspect a 4-5 \AA\ window around the line center. We then carefully select the continuum/pseudocontinuum regions around the line-of-interest, making sure to exclude all other absorption features. From the selected continuum points, we perform an iterative sigma-clipping, 2$\sigma$ then 1.5$\sigma$, to keep only those continuum/pseudocontinuum points with the best fit between the base model and the observed spectrum. For those remaining continuum/pseudocontinuum points, we perform a polynomial fit to the residuals R = O/M, where O is the observed flux and M is the interpolated model flux at each shifted, observed wavelength. Finally, we divide the observed spectrum by the polynomial fit to re-normalize and align continuum/pseudocontinuum levels of the observed spectrum to the models. See Figure \ref{pre_re_norm} for an example renormalization using an isolated $^{13}$CH line.

Unfortunately, even after a careful line-by-line renormalization, there are still regions of the observed spectrum that our models are not able to reproduce successfully. This may be due to spectral noise  and bad pixels in the observed spectrum,  insufficiencies in  model assumptions and  opacity calculations as well as atomic and molecular line lists incorporated in synthesizing spectra. Other factors such as NLTE  effects are typically negligible in solar type stars \citep{Nissen_Gustafsson_2018}. Magnetic field effects are also negligible when measuring CNO abundances \citep{Shchukina_2016} but may result in an underestimate of iron abundances \citep{Fabbian_2012}. Regardless, this means that we must remove lines from our analysis where the poor fit interferes with our ability to reliably determine the stellar $^{12}$C/$^{13}$C ratio. This leaves only three candidate lines for the HIRES spectrum and five lines for the ESPRESSO spectrum where both the continuum/pseudo-continuum and nearby atomic absorption lines of the model spectra align well with that of the observed spectrum. 

\subsection{Statistical Significance}
We remind the reader that our HIRES spectrum has a resolution R $\sim$ 45,000 and a median signal-to-noise ratio of 82 over the wavelength range of 3600-4400 \AA. Although this spectrum is fairly high S/N, the $^{13}$CH lines in this study are exceedingly weak. Therefore, we calculate the BIC factor (BIC = $\chi^2$ + n$_{dof}$*ln(N$_{data}$), where n$_{dof}$ is the number of degrees of freedom = 1, and N$_{data}$ is the number of data points used in the fit) for the selected $^{13}$CH lines to test the statistical significance of our detection. We determine $\Delta$BIC$_{160-BF}$ (BIC between the observed spectrum and the model with $^{12}$C/$^{13}$C = 160 minus BIC between the observed spectrum and the best-fit $^{12}$C/$^{13}$C model) for the $^{13}$CH lines in Table \ref{tab:crat}. This $\Delta$BIC$_{160-BF}$ tells us if the observed spectrum significantly favors low $^{13}$C-enrichment (or the $^{12}$C/$^{13}$C = 160 model) or higher $^{13}$C-enrichment (or the best fit $^{12}$C/$^{13}$C model). 
Table \ref{tab:crat} demonstrates a clear difference in the BIC factor, for each of our selected $^{13}$CH lines, for both the HIRES and ESPRESSO analysis. This means that both observed spectra favor models with solar to super-solar $^{13}$C enrichment.

\subsection{Deriving Individual Line $^{12}$C/$^{13}$C Ratios}\label{subsec: crats}
\begin{figure*}
\epsscale{1.00}
\plottwo{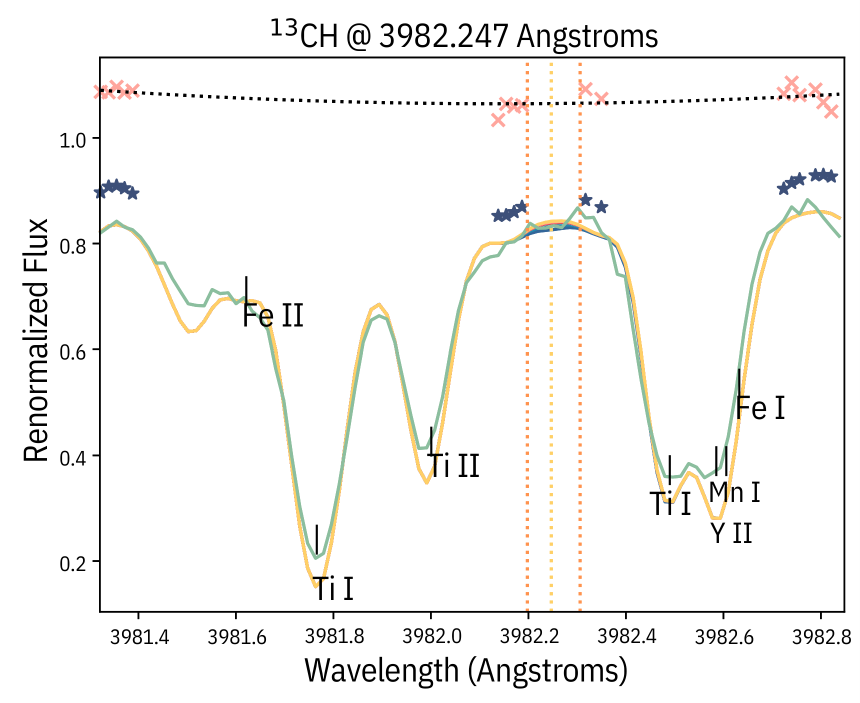}{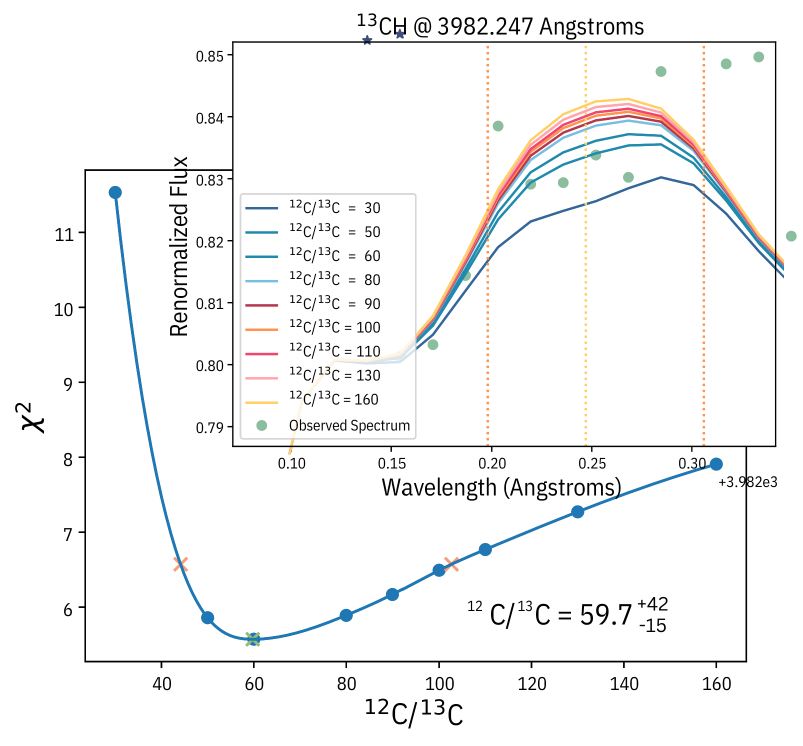}
\plottwo{figures/line13_renorm_final.pdf}{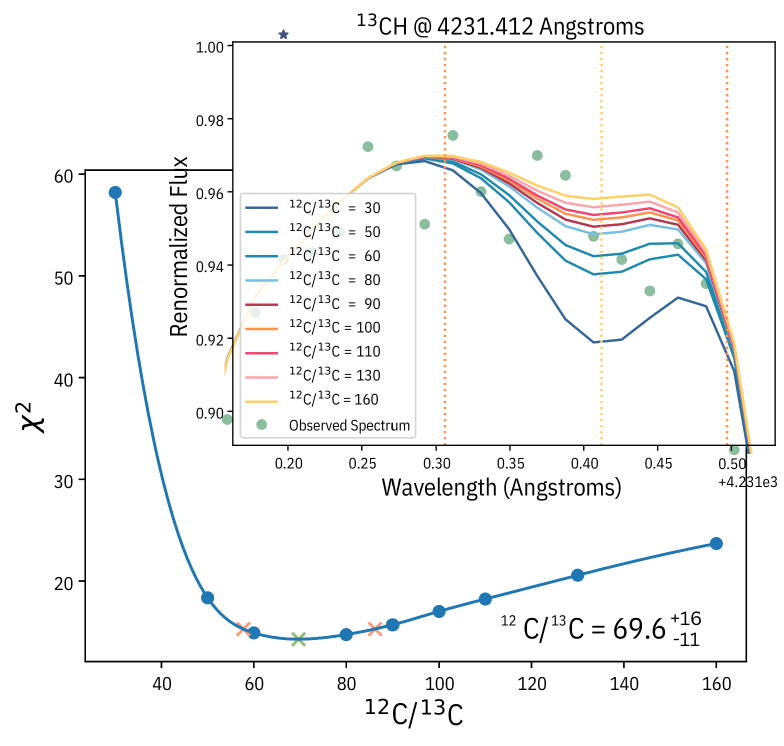}
\plottwo{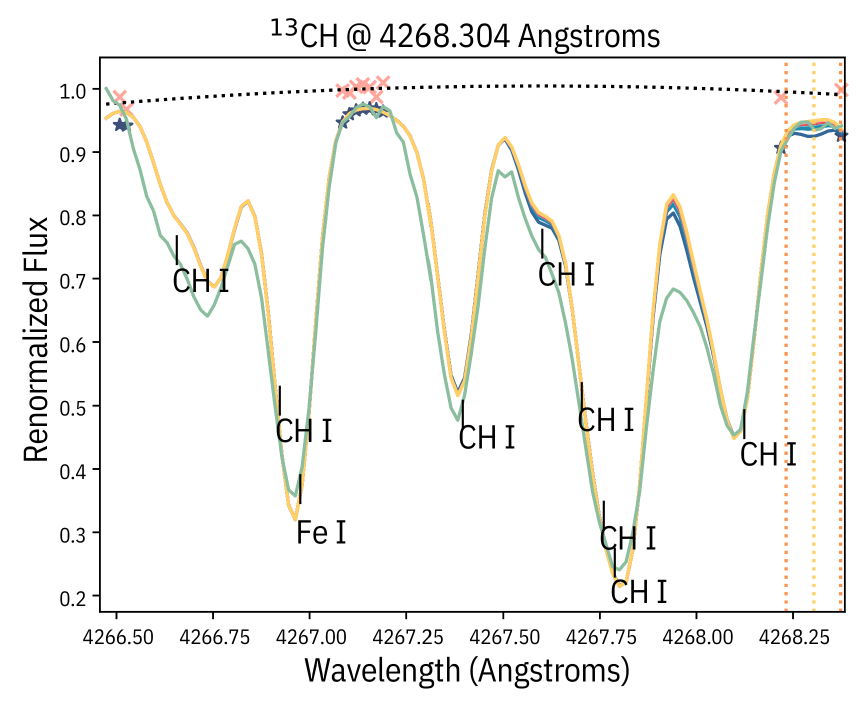}{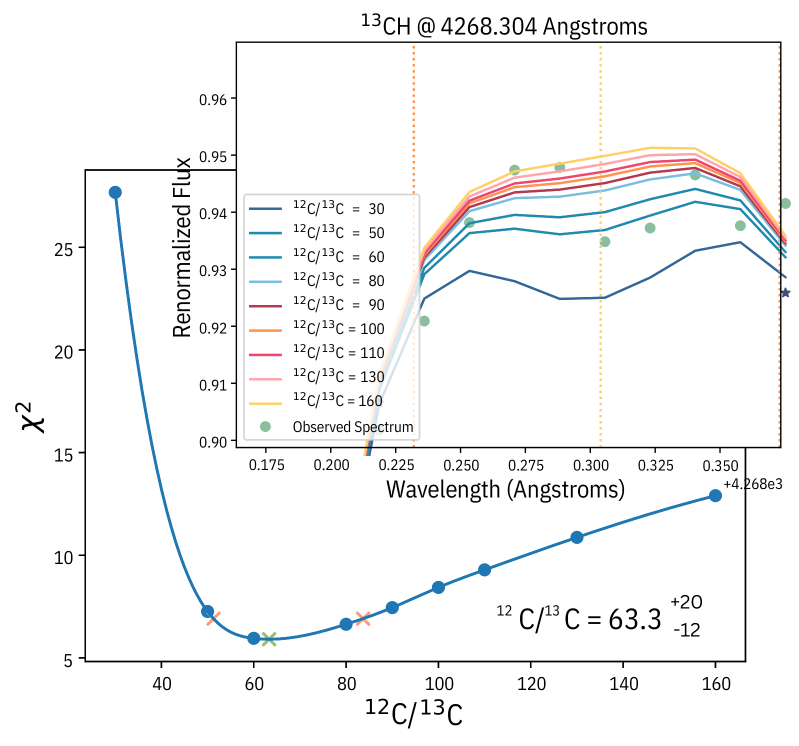}
\caption{Interpolating the stellar $^{12}$C/$^{13}$C ratio from our high-resolution and high signal-to-noise Keck/HIRES spectrum. On the left, we show our three selected $^{13}$CH lines, post-renormalization. We have labeled, in black, some nearby atomic and molecular lines that help us determine the best-fit base model spectrum. Refer to Table \ref{tab:line_list} for the complete line list. On the right, we zoom in on the $\chi^2$ fit windows to better see the separation between model spectra with high and low $^{12}$C/$^{13}$C ratios. We then calculate $\chi^2$ values between the model spectra and the observed spectrum for each target line, and fit the points with a cubic spline. The minimum of the spline represents the best fit $^{12}$C/$^{13}$C for the individual $^{13}$CH line.}
\label{chi_sqr}
\end{figure*}

We measure the $^{12}$C/$^{13}$C ratio in WASP-77 A following a similar process to that described in \cite{Crossfield_2019, Coria_2023}. We calculate $\chi^2$ between the re-normalized observed spectrum and each model spectrum, for each line over a wavelength range (typically 0.1-0.3 Angstroms) previously determined during the continuum selection process. For each set of $\chi^2$ values, we use a cubic spline fit to interpolate the minimum $\chi^2$ which corresponds to the best-fit $^{12}$C/$^{13}$C abundance. We infer 1$\sigma$ confidence intervals using the region where $\Delta\chi^2 \leq 1$ \citep{Avni_1976}. This gives us an estimate of the stellar $^{12}$C/$^{13}$C ratio for each selected $^{13}$CH line. Figures \ref{chi_sqr} - \ref{ESP} show the renormalized spectra, a zoom-in on the $^{13}$CH line, the $\chi^2$ minimization, and the derived $^{12}$C/$^{13}$C ratio for our selected lines in the HIRES and ESPRESSO spectra. Table \ref{tab:crat} contains more details on the $\chi^2$ minimization including the $\chi^2$ fit window, the number of points fit, the $\chi^2$ at the minimum associated with the best fit $^{12}$C/$^{13}$C ratio for each line. We obtain three individual $^{12}$C/$^{13}$C from the HIRES analysis and five from the ESPRESSO analysis, respectively. These values range from $^{12}$C/$^{13}$C = 47 - 75. Please refer to Table \ref{tab:crat} for the list of $^{12}$C/$^{13}$C ratios and their statistical uncertainties.

\subsection{Systematic Uncertainties}\label{subsec: sys_uncert}
Prior to determining the final $^{12}$C/$^{13}$C ratio for WASP-77A, we must consider the systematic uncertainties involved, particularly those associated with our choice of fundamental stellar parameters and elemental abundances. To test how the selection of fundamental stellar parameters impacts the derived $^{12}$C/$^{13}$C ratios, we generate several more model grids, this time varying effective temperature, surface gravity, bulk stellar metallicity, and individual C, O, Ti, and Fe abundances.  We find that changing individual elemental abundances (of species in close proximity to selected $^{13}$CH lines: [Fe/H], [Ti/H], [C/H], [O/H]) over the range reported in the literature values (+0.1, -0.2 dex), hardly change the model fit and the derived $^{12}$C/$^{13}$C ratios at all. However, effective temperature, surface gravity, bulk stellar metallicity do have a noticeable effect on the derived $^{12}$C/$^{13}$C ratios.

To determine systematic uncertainties of the derived $^{12}$C/$^{13}$C ratio due to changes in effective temperature, we vary the stellar effective temperature over span comparable various literature measurements, in steps of 25 K around our chosen value of 5545 K, while holding log g and metallicity constant. We find that this plausible range of effective temperature values produces $^{12}$C/$^{13}$C ratios ranging from 90-110 for the $^{13}$CH line at 4231.412 \AA. However, the changes in $^{12}$C/$^{13}$C due to the step change in effective temperature ($\sigma_{T} = 3$) are much smaller than the statistical uncertainties on our measurement of $^{12}$C/$^{13}$C derived from the HIRES spectrum ($\sigma = 17$). Thus, we are not limited by the uncertainties on the stellar parameters in this case. For the ESPRESSO spectrum, where the data is much more precise, the changes in $^{12}$C/$^{13}$C due to the step change in effective temperature ($\sigma_{T} = 4$) are larger than the statistical uncertainties on our measurement of $^{12}$C/$^{13}$C ($\sigma = 1$). In this case, we are somewhat limited by uncertainties on the stellar parameters.

We repeat the process to determine systematic uncertainties of the derived $^{12}$C/$^{13}$C ratio due to changes in surface gravity. We now vary the stellar surface gravity in steps of 0.05 around our chosen value of 4.45, while holding effective temperature and metallicity constant. We find systematic uncertainties in the derived $^{12}$C/$^{13}$C due to the change in log g  to be: $\sigma_{G} = 2$ for the HIRES analysis and $\sigma_{G} = 3$ for the ESPRESSO analysis. Again, these uncertainties limit the precision of the $^{12}$C/$^{13}$C ratio derived from the higher-quality ESPRESSO spectrum, but not the HIRES spectrum.

Finally, to determine systematic uncertainties of the derived $^{12}$C/$^{13}$C ratio due to changes in metallicity (Z), we deviate Z in steps of 0.1 dex. These steps are larger than the uncertainties of stellar metallicity reported in the literature ( $< 0.05$ dex), but we did this to inspect if non-solar metallicities would better reproduce spectral lines such as Ti, Mn, and CH. So, we vary the stellar metallicity in steps of 0.1 dex around our chosen value of +0.00 dex, while holding effective temperature and log g constant. We find that this plausible range of Z values produces a much broader range of $^{12}$C/$^{13}$C ratios ranging from 80-140. However, none of the reasonable Z values produce a model spectrum that significantly improves the fit to nearby spectral lines such as Ti, Mn, and CH. In the ESPRESSO spectrum, the models derived using Z = +0.00 dex fit much better than they do the HIRES spectrum.

We find that the derived $^{12}$C/$^{13}$C changes by $\sigma_{Z} = 7$ for the HIRES analysis and $\sigma_{Z} = 6$ for the ESPRESSO spectrum with each step of $\pm 0.1$ dex in bulk metallicity Z. These $\sigma_{Z}$ are overestimates, however, since the actual uncertainties on the stellar metallicity are less than half of our chosen step size. Therefore, we report final systematic uncertainties in the derived $^{12}$C/$^{13}$C ratio due to changes in stellar metallicity as $\sigma_{Z} = 3.5$ for the HIRES analysis and $\sigma_{Z} = 3$ for the ESPRESSO analysis.

Finally, we calculate the quadrature sum of $\sigma_{T}$, $\sigma_{G}$, and $\sigma_{Z}$ as the final systematic uncertainties in the $^{12}$C/$^{13}$C ratio: $\sigma_{sys} = 5$ for the HIRES analysis and $\sigma_{sys} = 6$ for the ESPRESSO analysis.

\subsection{Deriving the Stellar $^{12}$C/$^{13}$C Ratio}\label{subsec: crats}
Because the target $^{13}$CH spectral lines have low statistical significance and are barely visible by eye when considered individually, we derive the final stellar $^{12}$C/$^{13}$C ratio from these individual estimates collectively using both statistical uncertainties and the systematic uncertainties discussed above. We determine a final, weighted average of $^{12}$C/$^{13}$C = $66 \pm 18$ for WASP-77 A with our HIRES spectrum analysis, and a consistent, but more precise $^{12}$C/$^{13}$C = $51 \pm 6$ from our ESPRESSO spectrum analysis. We use the methods detailed in \cite{Barlow_2004} Section 2.8, "Model 2'' to calculate a weighted mean using asymmetric uncertainties.

Contrary to frequent assumptions, we do not find a solar $^{12}$C/$^{13}$C ratio for WASP-77A, but rather a sub-solar and even near-ISM $^{12}$C/$^{13}$C ratio (solar $^{12}$C/$^{13}$C = $91.4 \pm 1.3$, ISM $^{12}$C/$^{13}$C $\approx$ 68) \citep{Ayres_2013}. Compared to its host star, WASP-77A b appears slightly enriched in $^{13}$C with $^{12}$C/$^{13}$C = $26.4 \pm 16.2$ \citep{Line_2021}. We further discuss what these new abundance ratios may mean for WASP-77A b's formation in the following section.

\begin{figure*}
\epsscale{1.00}
\plottwo{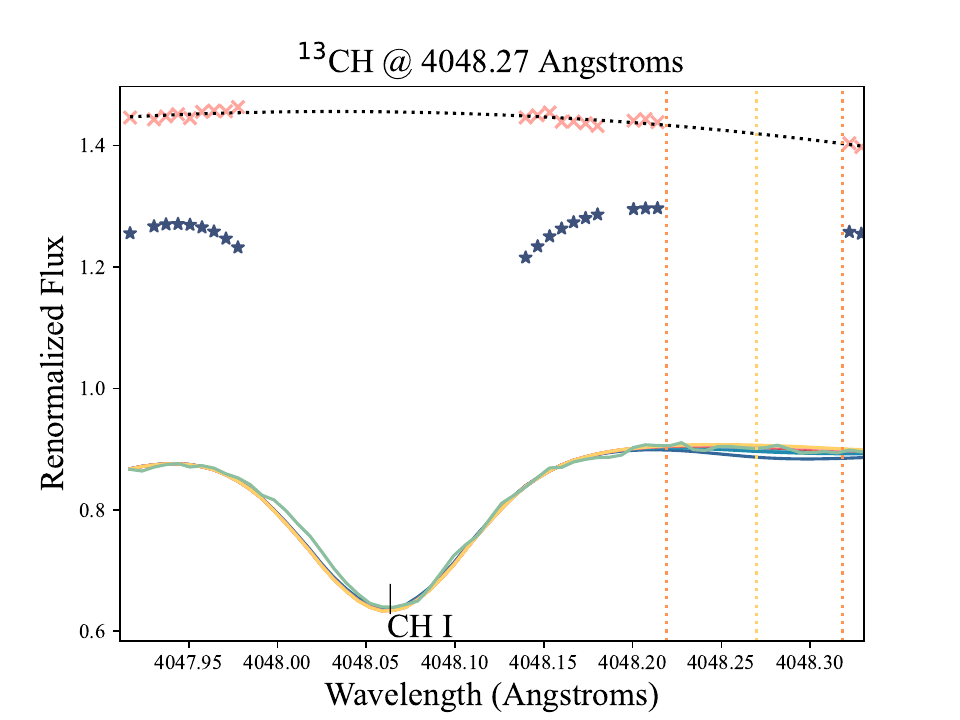}{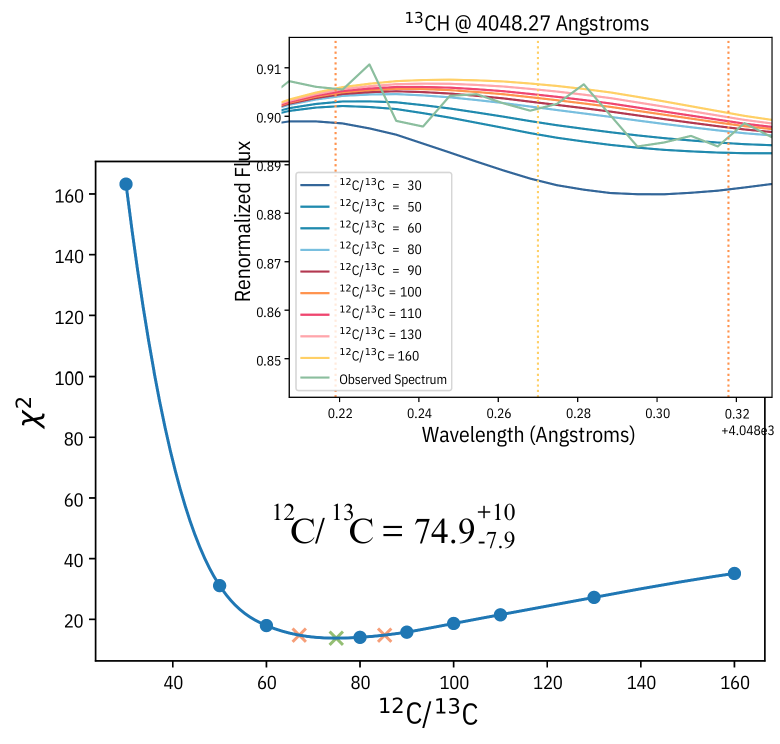}
\plottwo{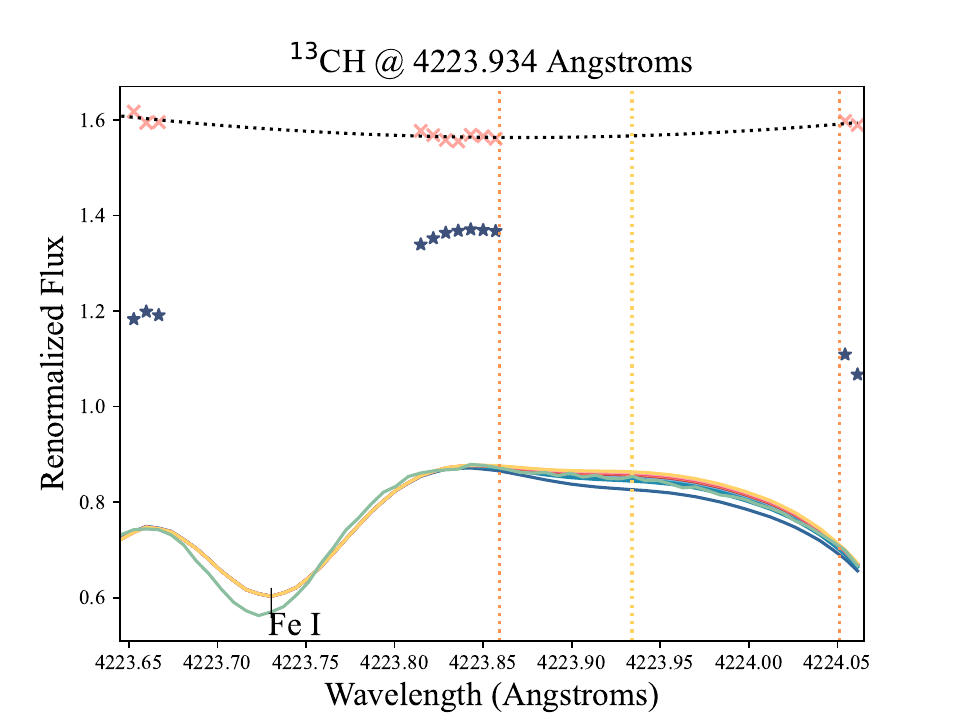}{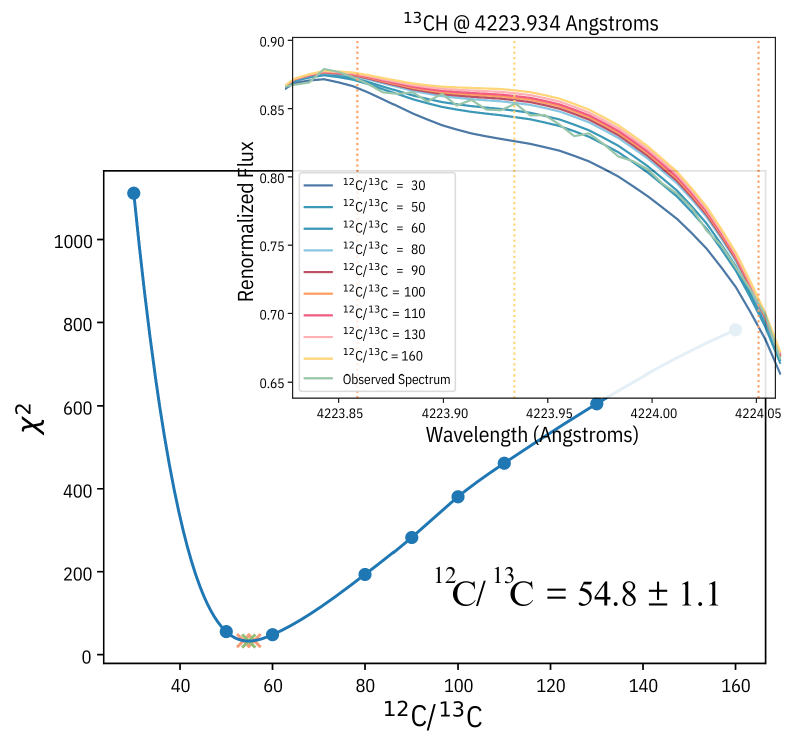}
\plottwo{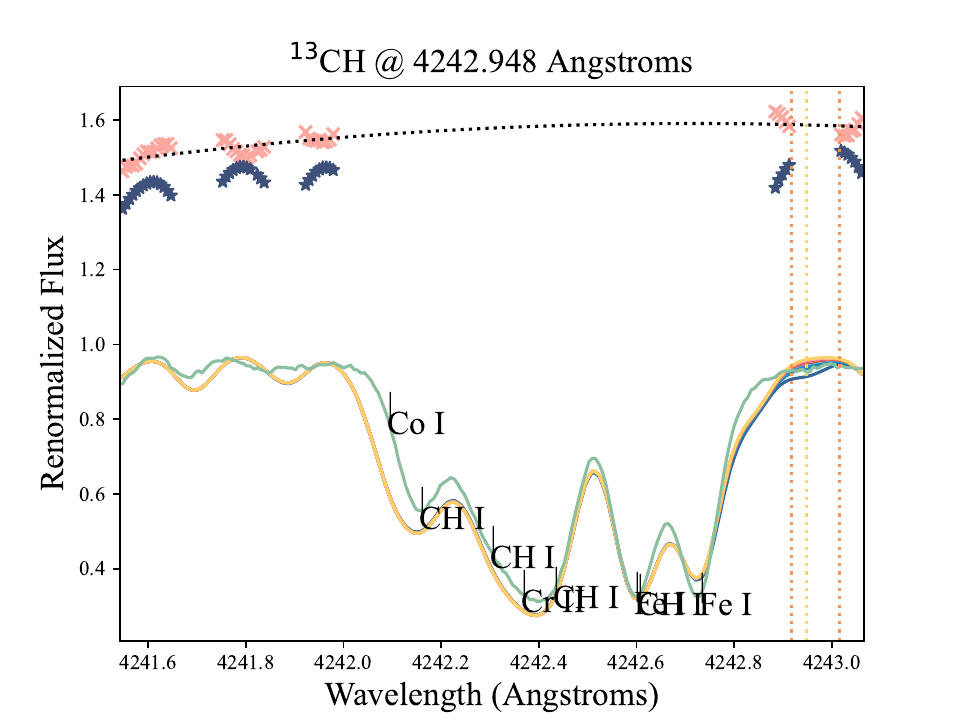}{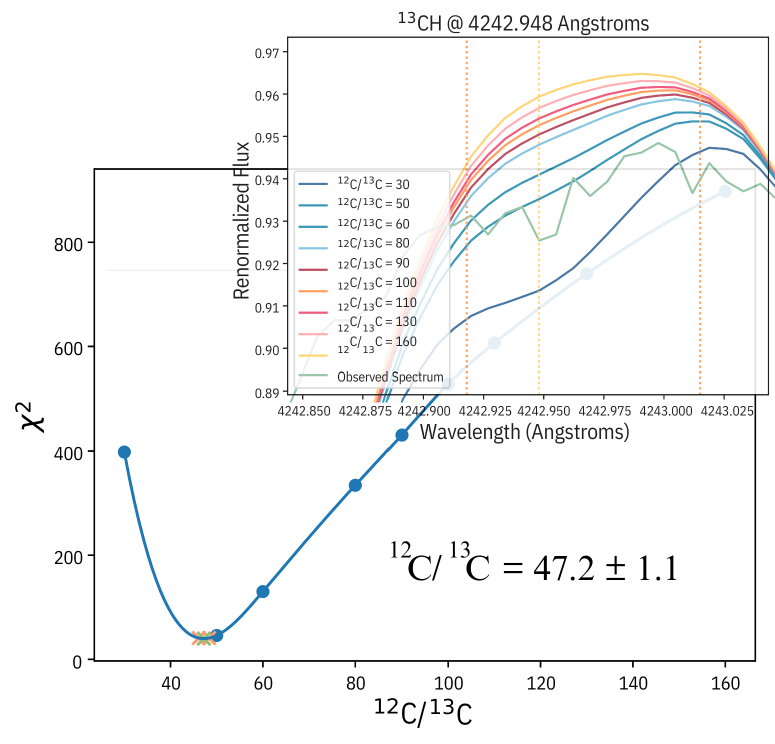}
\caption{Interpolating the stellar $^{12}$C/$^{13}$C ratio from the publicly available, high-resolution, and high signal-to-noise VLT/ESPRESSO spectrum. Similar to Figure \ref{chi_sqr}, we show our selected $^{13}$CH lines, post-renormalization on the left. On the right, we show a zoom-in on the $\chi^2$ fit windows and the $\chi^2$ minimization used to determine $^{12}$C/$^{13}$C for individual $^{13}$CH lines.}
\label{ESP}
\end{figure*}

\begin{figure*}
\epsscale{1.00}
\plottwo{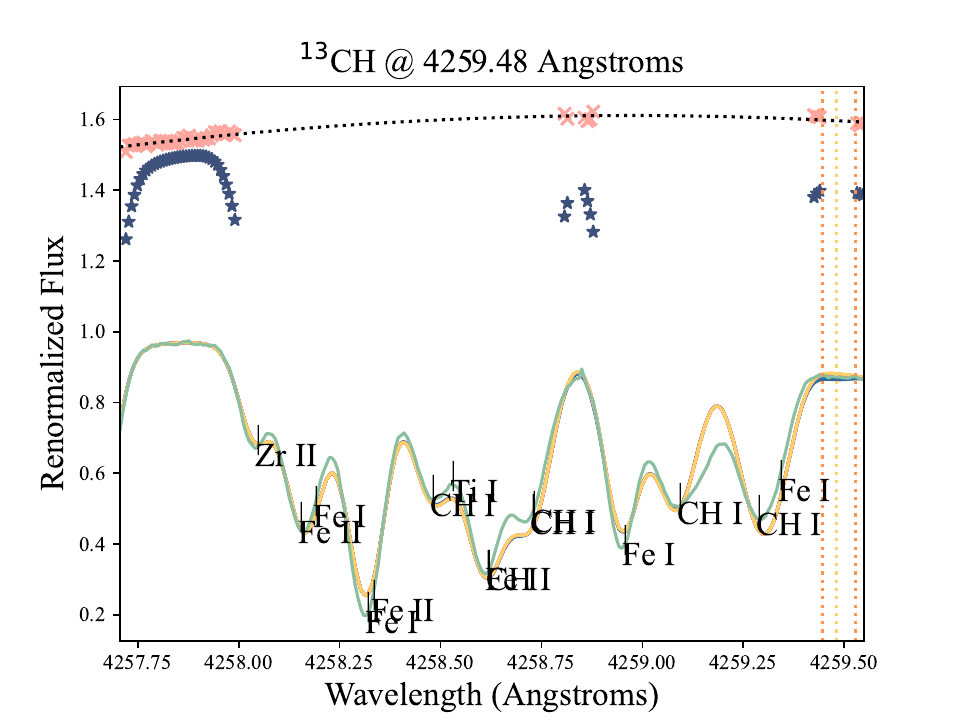}{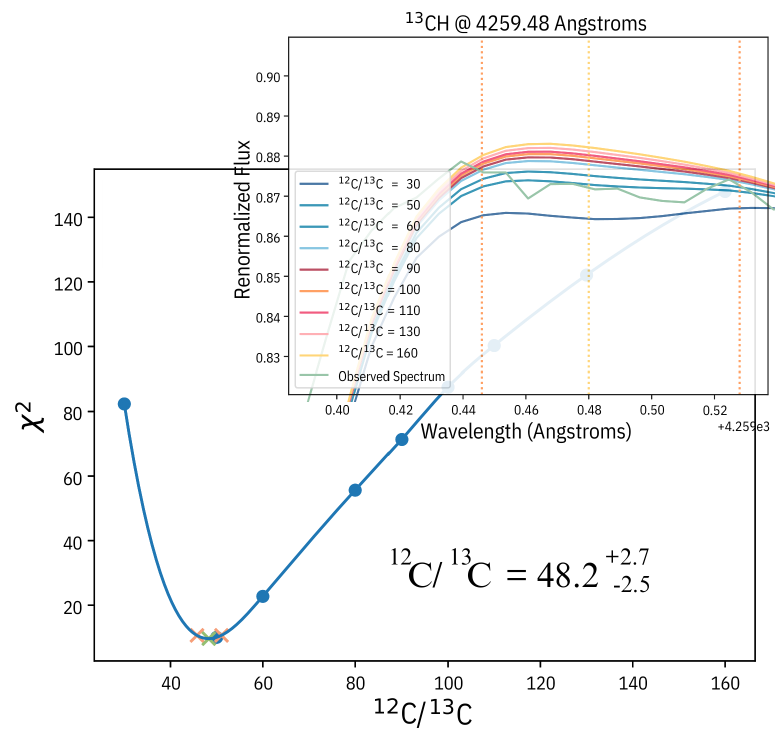}
\plottwo{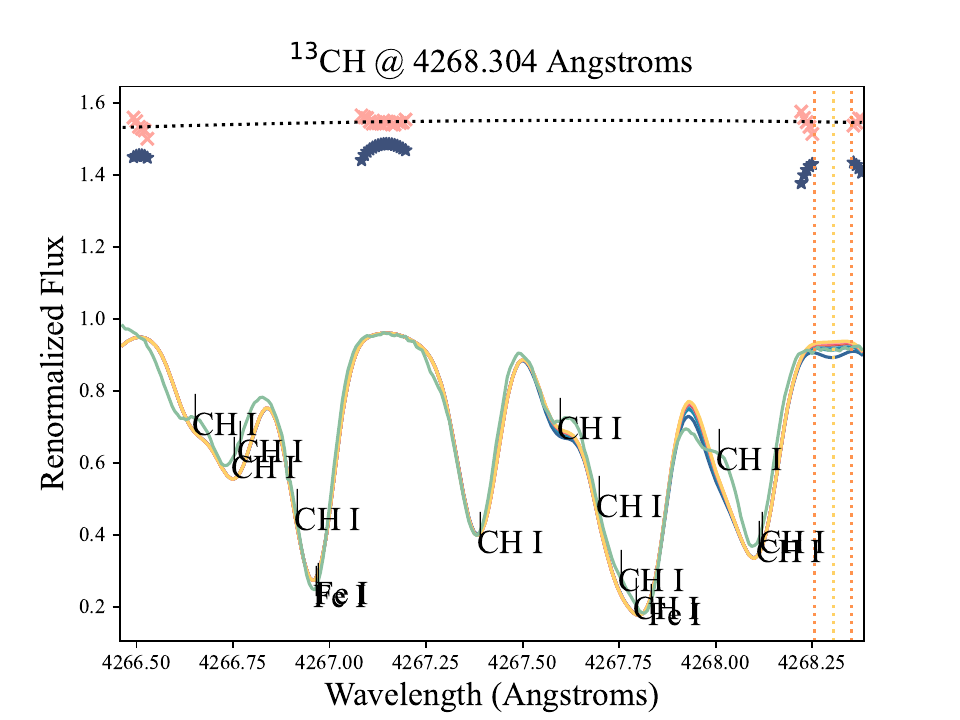}{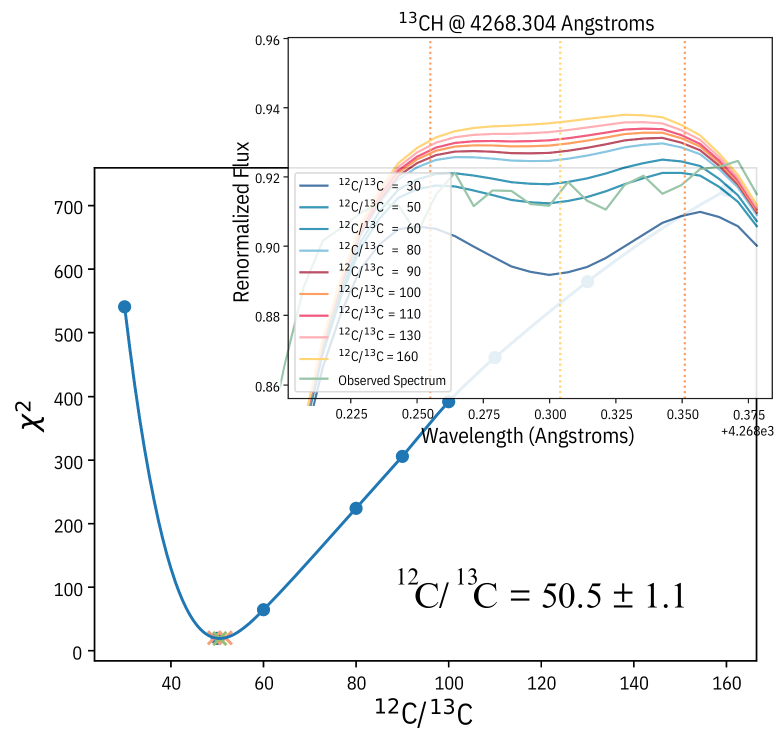}
\caption{Interpolating the stellar $^{12}$C/$^{13}$C ratio from the publicly available, high-resolution, and high signal-to-noise VLT/ESPRESSO spectrum. Similar to Figure \ref{chi_sqr}, we show our selected $^{13}$CH lines, post-renormalization on the left. On the right, we show a zoom-in on the $\chi^2$ fit windows and the $\chi^2$ minimization used to determine $^{12}$C/$^{13}$C for individual $^{13}$CH lines.}
\label{ESP}
\end{figure*}

\begin{deluxetable*}{ccccc}
\tablecaption{WASP-77 A (Host) vs. WASP-77 Ab (Planet) Abundances}
\tablewidth{0pt}
\tablehead{
\colhead{Host} & \colhead{Keck/HIRES} & \colhead{ARC/ARCES} & \colhead{MPG/ESO/FEROS} & \colhead{**Adopted} \\
\colhead{Parameter} & \colhead{\cite{Polanski_2022}} & \colhead{\cite{Reggiani_2022}} & \colhead{\cite{Kolecki_2022}} & \colhead{Parameters}
}
\startdata
T$_{eff}$ (K) & $5569 \pm 77$ & $5525 \pm 25$ & $5660 \pm 60$ & 5545 \\
log g & $4.45 \pm 0.09$ & $4.44 \pm 0.02$ & $4.49 \pm 0.03$ & 4.45 \\
$[Fe/H]$ & $0.01 \pm 0.03$ & $-0.05 \pm 0.02$ & $-0.15 \pm 0.06$ & +0.00 \\
$[C/H]$ & $-0.02 \pm 0.05$ & $0.10 \pm 0.09$ & $-0.04 \pm 0.04$ & +0.00 \\
$[O/H]$ & $0.06 \pm 0.07$ & $0.23 \pm 0.02$ & $-0.04 \pm 0.04$ & +0.00 \\
C/O & $0.46 \pm 0.09$ & $0.44 \pm 0.07$ & $0.59 \pm 0.08$ & - \\
\hline
\hline
\colhead{Planet} & \colhead{Gemini/IGRINS} & \colhead{HST/WFC3} & \colhead{JWST/NIRSpec} & \colhead{} \\
\colhead{Parameter} & \colhead{\cite{Line_2021}} & \colhead{\cite{Mansfield_2022}} & \colhead{\cite{August_2023}} & \colhead{}\\
\hline
[M/H] & $-0.48 \pm 0.15$ & $0.43 \pm 0.36$ & $-0.91 \pm 0.24$\\
$[C/H]$ & $-0.46 \pm 0.17$ & - & -\\
$[O/H]$ & $-0.49 \pm 0.17$ & - & - \\
C/O & $0.59 \pm 0.08$ & $< 0.78$ & $0.36 \pm 0.10$ \\
$^{12}$C/$^{13}$C & $26.4 \pm 16.2$ & - & - \\
\enddata

\tablecomments{Parameters and system-wide abundance inventory for host star WASP-77A and its hot Jupiter companion WASP-77A b. **This Paper. We derive the stellar $^{12}$C/$^{13}$C ratio for WASP-77A using the same HIRES spectrum as \cite{Polanski_2022} and a publicly available VLT/ESPRESSO spectrum.}
\label{tab:abundances}
\end{deluxetable*}

\clearpage

\startlongtable
\begin{deluxetable*}{cccc}
\tablecaption{Line List}
\tablewidth{0pt}
\tablehead{
\colhead{Species} & \colhead{Line Center} & \colhead{Equivalent Width} & \colhead{Renormalization}\\
\colhead{} & \colhead{(\AA)} & \colhead{(\AA)} & \colhead{Window (\AA)}
}
\startdata
$^{13}$CH* & 3982.247 & 1.60 & 3981.321 - 3982.848\\
Fe I & 3981.512 & 32.735 & - \\
Fe II & 3981.612 & 14.871 & - \\
Ti I & 3981.762 & 111.763 & - \\
Fe I & 3981.771 & 111.74 & - \\
Ti II & 3981.991 & 69.622 & - \\
Ti I & 3982.481 & 65.561 & - \\
Mn I & 3982.576 & 39.202 & - \\
Y II & 3982.592 & 81.178 & - \\
Fe I & 3982.624 & 17.647 & - \\
\hline
$^{13}$CH* & 4048.270 & - & 4047.913 - 4048.332\\
CH I & 4048.062 & 30.16 & - \\
\hline
$^{13}$CH* & 4223.934 & 1.74 & 4223.645 - 4224.065\\
Fe I & 4223.729 & 27.096 & - \\
\hline
$^{13}$CH* & 4231.412 & 1.84 & 4230.701 - 4231.860\\
Ni I & 4231.032 & 46.739 & - \\
Fe I & 4231.597 & 12.83 & - \\
CH I & 4231.000 &  67.704 & - \\
CH I & 4231.035 & 69.041 & - \\
CH I & 4231.592 & 41.903 & - \\
CH I & 4231.680 & 43.164 & - \\
\hline
$^{13}$CH* & 4242.948 & 2.13 & 4241.542 - 4243.066\\
Co I & 4242.089 & 15.963 & - \\
Cr II & 4242.364 & 52.167 & - \\
Fe I & 4242.595 & 49.534 & - \\
Fe I & 4242.729 & 60.709 & - \\
CH I & 4242.156 & 44.291 & - \\
CH I & 4242.300 & 46.168 & - \\
CH I & 4242.430 & 69.514 & - \\
CH I & 4242.603 & 72.165 & - \\
\hline
$^{13}$CH* & 4259.480 & - & 4257.706 - 4259.549\\
Zr II & 4258.041 & 23.588 & - \\
Fe II & 4258.148 & 43.764 & - \\
Fe I & 4258.186 & 29.79 & - \\
Fe I & 4258.315 & 85.959 & - \\
Fe II & 4258.328 & 16.086 & - \\
Ti I & 4258.524 & 11.828 & - \\
Fe I & 4258.611 & 68.699 & - \\
Fe I & 4258.951 & 52.447 & - \\
Fe I & 4259.336 & 22.261 & - \\
CH I & 4258.474 & 45.463 & - \\
CH I & 4258.613 & 24.176 & - \\
CH I & 4258.722 & 48.086 & - \\
CH I & 4258.725 & 25.54 & - \\
CH I & 4259.087 & 45.468 & - \\
CH I & 4259.282 & 48.09 & - \\
\hline
$^{13}$CH* & 4268.304 & 2.14 & 4266.457 - 4268.384\\
Fe I & 4266.958 & 14.204 & - \\
Fe I & 4266.964 & 79.921 & - \\
Fe I & 4267.826 & 96.387 & - \\
CH I & 4266.644 & 26.397 & - \\
CH I & 4266.745 & 25.4 & - \\
CH I & 4266.762 & 26.443 & - \\
CH I & 4266.910 & 26.815 & - \\
CH I & 4267.383 & 69.777 & - \\
CH I & 4267.589 & 25.404 & - \\
CH I & 4267.692 & 26.819 & - \\
CH I & 4267.748 & 72.441 & - \\
CH I & 4267.787 & 69.781 & - \\
CH I & 4268.002 & 30.661 & - \\
CH I & 4268.105 & 72.445 & - \\
CH I & 4268.113 & 30.14 & - \\
\enddata
\tablecomments{We include species, line center, and  equivalent width for atomic and molecular lines within the renormalization window for each of our $^{13}$CH lines of interest (starred). CH line lists are from \cite{Masseron_2014}. Atomic lines are from the Vienna Atomic Line Database \citep{Ryabchikova_2015}. Equivalent widths are calculated using TurboSpectrum \citep{Plez_2012}. Lines without equivalent widths were identified by eye.}
\label{tab:line_list}
\end{deluxetable*}

\begin{deluxetable*}{cccccccc}
\tablecaption{$\chi^2$ Minimization and Individual $^{12}$C/$^{13}$C Ratios}
\tablewidth{0pt}
\tablehead{
\colhead{Species} & \colhead{Line Center} & \colhead{$\chi^2$ Fit Window} & \colhead{\# of Points Fit} & \colhead{$\chi^2$ at Minimum} & \colhead{$\Delta$BIC$_{160 - Best}$} & \colhead{$^{12}$C/$^{13}$C} & \colhead{$^{12}$C/$^{13}$C} \\
\colhead{} & \colhead{\AA} & \colhead{\AA} & \colhead{H, E} & \colhead{H, E} & \colhead{H, E} & \colhead{H}& \colhead{E}
}
\startdata
$^{13}$CH & 3982.247 & 3982.198 - 3982.306 & 7, - & 5.4, - & $\Delta$BIC$_{160-50}$ = 19.775, - & 59.7$^{+42}_{-15}$ & -\\
$^{13}$CH & 4048.270 & 4048.219 - 4048.318 & 5, 15 & -, 13.7 & -, $\Delta$BIC$_{160-80}$ = 21.050 & - & 74.9$^{+10}_{-7.9}$\\
$^{13}$CH & 4223.934 & 4223.859 - 4224.051 & 11, 27 & -, 32.7 & -, $\Delta$BIC$_{160-60}$ = 733.981 & - & 54.8$^{+1.1}_{-1.1}$\\
$^{13}$CH & 4231.412 & 4231.306 - 4231.497 & 10, 21 & 14.6, - & $\Delta$BIC$_{160-100}$ = 6.691, - & 69.6$^{+16}_{-11}$ & -
\\
$^{13}$CH & 4242.948 & 4242.880 - 4243.015 & -, 14 & -, 39.8 & -, $\Delta$BIC$_{160-50}$ = 852.014 & - & 47.2$^{+1.1}_{-1.0}$\\
$^{13}$CH & 4259.480 & 4259.399 - 4259.528 & 7, 12 & -, 9.64 & -, $\Delta$BIC$_{160-50}$ = 137.960 & - & 48.2$^{+2.7}_{-2.5}$\\
$^{13}$CH & 4268.304 & 4268.232 - 4268.373 & 8, 14 & 5.9, 19.2 & $\Delta$BIC$_{160-50}$ = 39.618, 704.753 & 63.3$^{+20}_{-12}$ & 50.5$^{+1.1}_{-1.1}$\\ 
\enddata
\tablecomments{The first two columns show the location of our selected $^{13}$CH lines. Columns 3 describes the location of the $\chi^2$ fit window including the number data points fit (Column 4), the calculated $\chi^2$ minimum (Column 5), and the $\Delta$BIC$_{160 - Best}$ (Column 6) associated with the best fit $^{12}$C/$^{13}$C ratios for the HIRES (H) and ESPRESSO (E) spectra (Columns 7 and 8, respectively). $\Delta$BIC$_{160 - Best}$ is calculated as the BIC difference between the lowest $^{13}$C-enriched model with $^{12}$C/$^{13}$C = 160, and the best-fit model.}
\label{tab:crat}
\end{deluxetable*}

\section{Discussion} \label{sec:discussion}
We now see slight deviations from solar abundances in WASP-77A's photosphere. The weighted average of the three stellar abundance values from \citep{Polanski_2022, Reggiani_2022, Kolecki_2022} are [C/H] = $-0.02 \pm 0.03$, [O/H] = $0.17 \pm 0.02$, C/O = $0.49 \pm 0.05$, and [Fe/H] = $-0.05 \pm 0.02$. WASP-77A potentially has [C/H] consistent with solar, elevated [O/H], sub-solar C/O (solar C/O = $0.59 \pm 0.08$ \citep{Asplund_2021}), and slightly sub-solar metallicity. Additionally, we now see a potential difference between the solar $^{12}$C/$^{13}$C ratio ($91.4 \pm 1.3$) and our measurement of WASP-77A's $^{12}$C/$^{13}$C ratio ($51 \pm 6$).

Now, by comparing host star WASP-77A's carbon, oxygen, and metallicity values to those of its hot Jupiter companion, we notice WASP-77A b's sub-stellar [C/H] = $-0.46 \pm 0.17$, significant oxygen depletion with [O/H] = $-0.49 \pm 14$, C/O = $0.50 \pm 0.06$ consistent with its host star, and sub-stellar metallicity with [M/H] = $-0.60 \pm 0.13$.

\subsection{WASP-77A b's Formation Location}
Hot Jupiter systems are of particular interest for abundance studies because of their amenability for observation, but also because of their formation histories. There is only sufficient material for in-situ hot Jupiter formation in large protoplanetary disks where there are also likely to be multiple cold Jupiters in the outer disk \citep{Batygin_2016}. This does not appear to be the case for WASP-77A b. Thus, we expect non-stellar abundances in hot-Jupiter atmospheres, characteristic of the planet's formation location and accretion history elsewhere in the disk. The following formation scenario for WASP-77 A b relies on the volatile abundance inventory for the WASP-77 A system (Table \ref{tab:abundances}: including both host star and planetary C/H, O/H, C/O, and $^{12}$C/$^{13}$C ratios) and its usage in determining formation and migration diagnostics as discussed in Sections \ref{elemental_ab_diagnostics} and \ref{isotopic_ab_diagnostics}.

A sub-stellar C/O ratio would indicate radial migration where the planet accumulates O-rich and C-poor material, however, WASP-77A b's C/O ratios are mostly consistent with its host star \citep{Line_2021, August_2023}. While the similar C/O ratios alone may be indicative of formation close to the star (within the H$_2$O snowline at 5 AU), it is critical to take into account individual carbon and oxygen abundances as well. The sub-stellar carbon and oxygen abundances measured in WASP-77A b's atmosphere are a signature of formation in the outer disk, beyond the H$_2$O and CO$_2$ snowlines (located at around 5 and 10 AU respectively), where most carbon and oxygen is trapped in grains, leaving the planet-feeding gas depleted in volatiles \citep{Bosman_2021}. Planets forming in this outer region of the disk are expected to have sub-stellar C/H and O/H. In fact, the sub-solar C/H, O/H, and solar C/O ratios of WASP-77A b are well reproduced by planet formation and accretion models which predict the planet's formation beyond the CO$_2$ evaporation front \citep{Bitsch_2022}. Overall, WASP-77A b's sub-stellar atmospheric carbon and oxygen abundances and stellar C/O ratio are indicative of this planet's formation beyond the host's H$_2$O and CO$_2$ snowlines located at $\sim$ 5AU and 10 AU respectively \citep{Oberg_2011, Reggiani_2022, Bitsch_2022}.

Though elemental abundances alone constrain WASP-77A b's formation location to be beyond both the H$_2$O and CO$_2$ snowlines, or somewhere at a separation greater than 10 AU, our goal was to make a complementary measurement of the stellar $^{12}$C/$^{13}$C ratio to determine whether (1) WASP-77A b's $^{13}$C enrichment was inherited from the host star, placing the formation location between the CO$_2$ snowline at 10 AU and the CO snowline at 20 AU or (2) whether WASP-77A b's $^{13}$C enrichment was due to the accretion of $^{13}$C-rich material found beyond the CO snowline at 20 AU. Our additional finding that WASP-77A has a sub-solar $^{12}$C/$^{13}$C ratio of $51 \pm 6$ shows that the hot Jupiter WASP-77A b may indeed have $^{13}$C enrichment ($^{12}$C/$^{13}$C = 10.2-42.6), but perhaps not as significant as when we were assuming a solar $^{12}$C/$^{13}$C for the host star. Again, protoplanetary disk chemistry models produce a broad range of $^{12}$C/$^{13}$C in CO as mid-plane height and radial distance from the star varies \citep{Zhang_2017, Nomura_2023, Bergin_2024, Yoshida_2024, Line_2021, Woods_Willacy_2009}. Although this host star and exoplanet $^{12}$C/$^{13}$C ratio comparison appears consistent with previous studies that place WASP-77A b's formation beyond the stellar H$_2$O and CO$_2$ snowlines located at 5 and 10 AU respectively, it is unclear whether the difference is considerable enough to suggest formation beyond the stellar CO snowline where significant $^{13}$C enrichment is expected occur \citep{Zhang_2021a, Bergin_2024}. Notably, WASP-77A b has a $^{12}$C/$^{13}$C ratio inconsistent with that of any planet in our own solar system, but this is perhaps unexpected as WASP-77A b's formation and accretion history is thought to be much different than that of our solar system planets. The $^{13}$C-enrichment found in WASP-77A b's atmosphere, relative to its host star, and the elemental carbon and oxygen abundances in the system disfavor an ``in situ" formation scenario nonetheless. This new piece of evidence corroborates our understanding of hot Jupiter formation from a dynamical standpoint. That is, hot-Jupiters likely did not form in situ, but rather migrated to their current positions from the outer disk. Refer to Section \ref{subsec:migration} for a brief dynamical explanation on how hot Jupiters form.

We emphasize that the interpretation of abundance ratios measured in giant exoplanet atmospheres relies on the abundance ratios measured in the host star's photosphere and on planet formation and accretion models. For a further analysis of the interplay between a planet's atmospheric carbon and oxygen abundances and different planet formation pathways in the context of the WASP-77 system, we refer the reader to \cite{Bitsch_2022}.

\subsection{Possible Migration Mechanisms}\label{subsec:migration}
There are several mechanisms for migration that could transform a warm/cool Jupiter into a short-period hot Jupiter. These include coplanar high-eccentricity migration \citep{Petrovich_2015b}, secular chaos \citep{Wu_Lithwick_2011}, and Lidov-Kozai cycling \citep{Wu_Murray_2003, Petrovich_2015a}. A recent study based on data from the California Legacy Survey \citep{Rosenthal_2021} argues that coplanar high-eccentricity migration is the most likely formation pathway for hot-Jupiter formation \citep{Zink_2023}. In this scenario, there are two cold gas giant planets that through the exchange of angular momentum and tidal interactions with the host star, culminate in a hot Jupiter. \cite{Zink_2023} finds that giant planet multiplicity is ubiquitous with an average of $1.3^{+1.0}_{-0.6}$ companions per hot Jupiter and $1.0 \pm 0.3$ companions per warm or cold Jupiter. This provides plenty of opportunities for hot Jupiter formation via coplanar high-eccentricity migration. Most notably, this study finds that hot Jupiter-class planets tend to have an outer companion with at least three times its mass. In the WASP-77 system, we do not see an additional, Jupiter-class companion with mass M $>$ 3 M$_J$ that may have driven this type of migration, but we do see a possible suspect in the secondary star, WASP-77B. This star is not well characterized, but studies find it is spectral type K5 with an effective temperature T$_{eff}$ = $4810 \pm 100$ K which places it a mass anywhere from 0.5-0.8 solar masses based on mass-effective-temperature relationships for stars on the main sequence \citep{Evans_2016}. This is at least three times more massive than WASP-77A b, and may be the cause of the planet's migration to its current location $\sim 0.02$ AU from its host star. 

\subsection{WASP-77B Contamination}
Although we have mostly discussed the primary star WASP-77A and its, companion planet WASP-77A b, there is a third member of the system: a K dwarf star WASP-77B  at a separation of approximately 3 arc-seconds \citep{Maxted_2013}. It could be argued that the peculiar abundances observed in WASP-77A b may be a result of chemical contamination from the secondary star WASP-77B. However, there is no clear avenue for determining elemental abundances for this star. There are only $\sim 4$ low signal-to-noise (Average S/N = 30) ESO/HARPS spectra of WASP-77B in the literature, no effective temperature, surface gravity, or metallicity, constraints and therefore no elemental abundances for this star in the literature. Nonetheless, such contamination seems unlikely as the seeing for our Keck/HIRES spectrum was 1.3'' and 0.45'' for the VLT/ESPRESSO spectrum.

\section{Conclusions}\label{sec:conclusions}
\subsection{Summary}
Using the James Webb Space Telescope, ground-based 8m- class telescopes, and ultra high-resolution spectroscopy, we are able to characterize the composition of exoplanet atmospheres like never before. However, it is necessary to properly contextualize these atmospheric abundance constraints by comparing them to their host star's abundances rather than to solar abundances. In the case of hot Jupiter WASP-77A b and its host star WASP-77A, metallicity, carbon, and oxygen abundances provide us an insight into the planet's origins beyond the H$_2$O and CO$_2$ snowlines \citep{Reggiani_2022, Bitsch_2022}. In this paper, we derive the $^{12}$C/$^{13}$C ratio for WASP-77A to complement a similar measurement in WASP-77A b made by \cite{Line_2021}. The sub-solar and near-ISM $^{12}$C/$^{13}$C ratio we find for WASP-77A ($51 \pm 6$) provides additional evidence for WASP-77A b's formation beyond the H$_2$O and CO$_2$ snowlines. 

\subsection{Prospects for Measuring Nitrogen and Oxygen Isotopologue Ratios in Cool Dwarf Stars}
With the advent of some astounding exoplanet abundance constraints \citep{Line_2021, Finnerty_2024, Zhang_2021a, Gandhi_2023}, we must continue building the catalogue of elemental and isotopic host star abundances as in \cite{Souto_2017, Souto_2018, Crossfield_2019, Hejazi_2023, Coria_2023, Hejazi_2024} with an emphasis on carbon, nitrogen, and oxygen. Stellar [N/H] and $^{14}$N/$^{15}$N abundances are best measured using optical CN \citep{Sneden_2014, Ryabchikova_2022} and possibly NH lines. While [C/H] and $^{12}$C/$^{13}$C may be measured using optical CH lines as we do here, they are preferentially measured for cooler K and M dwarfs in the near-infrared using CO isotopologue lines \citep{Goorvitch_1994, Gordon_2017}. In the near-infrared, we may also derive [O/H], $^{16}$O/$^{18}$O, and possibly $^{17}$O/$^{18}$O using CO, OH and H$_2$O isotopologue lines \citep{Goorvitch_1994, Goldman_1982, Barber_2006, Xuan_2024}.

\subsection{CNO Abundances in Brown Dwarfs}
Besides exoplanets, brown dwarf abundances also play an important role in understanding planet formation. Y dwarfs specifically are cool enough that we may see absorption from molecules that dissociate in the atmospheres of short period hot-Jupiters and wide-orbit young Jupiters. Recent studies are able to measure $^{12}$C/$^{13}$C \citep{Lew_2024, Hood_2024, Costes_2024}, $^{14}$N/$^{15}$N \citep{Barber_2006}, and $^{16}$O/$^{18}$O \citep{Zhang_2021b} in the atmospheres of cool brown dwarfs like 2M0355, WISE J1828, and 2M0415. CNO isotopologue ratios may provide chemical distinctions between brown dwarfs and super-Jupiters, or top-down vs. bottom-up formation models, as we start seeing ISM- or greater isotopologue ratios in brown dwarfs, but sub-stellar and sub-ISM isotopologue ratios in the super-Jupiters.

\subsection{CNO Abundances in the Context of Other Planetary Systems}
Jupiter-class exoplanets and their host stars provide the most practical systems with which to test how carbon, oxygen, and their respective isotopic abundance ratios change with radial distance from the host star. Specifically, the most amenable planetary systems to complementary abundance analyses are those in the WASP (Wide Angle Search for Planets) catalog \citep{Pollacco_2006} and the young, wide separation, directly imaged Jupiter-class planet systems. These planetary systems are routinely targeted by JWST and ground-based observatories for transmission and emission spectroscopy and provide the best avenue for the direct comparison of stellar and planetary carbon and oxygen abundances, both elemental and isotopic. These complementary abundance surveys do, however, come with several challenges exemplified by the WASP-121 and TYC 8998-760-1 systems. WASP-121 and its ultra-hot Jupiter companion WASP-121 b \citep{Delrez_2016} are well studied both in transmission and secondary eclipse observations \citep{Mikal-Evans_2018, Mikal-Evans_2019, Kovacs_2019, Sing_2019}. Though these observations may retrieve carbon and oxygen abundance constraints for the companion planet, complementary host star measurements are made much more difficult (and isotopic abundance measurements, impossible) by the star's fast rotation. The recently discovered TYC-8998-760-1 is a young, K2IV star host to two wide-separation Jupiter-class planets \citep{Bohn_2020}. As ESO/SPHERE direct imaging and spectroscopy prioritize observations of the planets \citep{Zhang_2021b}, the chemical composition of the host star remains obscure. Furthermore, the strong magnetic fields prevalent in these young stars make it difficult to model their spectra and determine fundamental effective temperature, surface gravity, and metallicity values-- let alone elemental abundances-- without some significant processing of the stellar spectrum \citep{Lopez_Valdivia_2021}. More work must be done to identify planetary systems where we may determine carbon, oxygen, and their respective isotopic abundances in both the host star and companion planet.

\begin{acknowledgments}
The authors of this work would like to thank the \dataset[California Planet Search]{https://exoplanets.caltech.edu/cps/} team for providing the HIRES spectrum (originally analyzed in \cite{Polanski_2022}) that was used in this analysis. We also thank the referee for an excellent discussion on the statistical significance of our measurements and the systematic uncertainties associated with our choice of model parameters which both greatly improved this paper. Finally, the authors would also like to acknowledge the support of National Science Foundation grant AST-2108686 and NASA ICAR grant A21-0406-S003. 
\end{acknowledgments}

%

\vspace{5mm}
\facilities{W. M. Keck Observatory}




\begin{thebibliography}{}
\bibitem[Adibekyan et al.(2012)]{Adibekyan_2012} Adibekyan, V. Z., Sousa, S. G., Santos, N. C., et al. 2012, A\&A, 545, A32, doi: 10.1051/0004-6361/201219401
\bibitem[Ali-Dib et al.(2017)]{Ali-Dib_2017} Ali-Dib, M. 2017, \mnras, 467, 2845, doi: 10.1093/mnras/stx260
\bibitem[Andrews et al.(2013)]{Andrews_2013} Andrews, S. M., Rosenfeld, K. A., Kraus, A. L.,andWilner, D. J. 2013, \apj, 771, 129, doi: 10.1088/0004-637X/771/2/129
\bibitem[Asplund et al.(2021)]{Asplund_2021} Asplund, M., Amarsi, A. M., and Grevesse, N. 2021, A\&A, 653, A141, doi: 10.1051/0004-6361/202140445
\bibitem[Atreya et al.(2016)]{Atreya_2016} Atreya, S. K., Crida, A., Guillot, T., et al. 2016, arXiv e-prints, arXiv:1606.04510, doi: 10.48550/arXiv.1606.04510
\bibitem[August et al.(2023)]{August_2023} August, P. C., Bean, J. L., Zhang, M., et al. 2023, \apjl. https://arxiv.org/abs/2305.07753
\bibitem[Avni et al.(1976)]{Avni_1976} Avni, Y. 1976, \apj, 210, 642, doi: 10.1086/154870
\bibitem[Ayres et al.(2013)]{Ayres_2013} Ayres, T. R., Lyons, J. R., Ludwig, H.-G., Caffau, E., and Wedemeyer-Bohm, S. 2013, \apj, 765, 46, doi: 10.1088/0004-637x/765/1/46
\bibitem[Barber et al.(2006)]{Barber_2006} Barber, R. J., Tennyson, J., Harris, G. J., and Tolchenov, R. N. 2006, \mnras, 368, 1087, doi: 10.1111/j.1365-2966.2006.10184.x
\bibitem[Barlow et al.(2004)]{Barlow_2004} Barlow, R. 2004, Asymmetric Errors. https://arxiv.org/abs/physics/0401042
\bibitem[Barrado et al.(2023)]{Barrado_2023} Barrado, D., Molli`ere, P., Patapis, P., et al. 2023, Nature, 624, 263, doi: 10.1038/s41586-023-06813-y
\bibitem[Batygin et al.(2016)]{Batygin_2016} Batygin, K., Bodenheimer, P. H., and Laughlin, G. P. 2016, \apj, 829, 114, doi: 10.3847/0004-637X/829/2/114
\bibitem[Bedell et al.(2018)]{Bedell_2018} Bedell, M., Bean, J. L., Mel´endez, J., et al. 2018, \apj, 865, 68, doi: 10.3847/1538-4357/aad908
\bibitem[Bergin et al.(2024)]{Bergin_2024} Bergin, E. A., Bosman, A., Teague, R., et al. 2024, arXiv, arXiv:2403.09739, doi: 10.48550/arXiv.2403.09739
\bibitem[Bitsch and Mah(2023)]{Bitsch_Mah_2023} Bitsch, B., and Mah, J. 2023, A\&A, 679, A11, doi: 10.1051/0004-6361/202347419
\bibitem[Bitsch et al.(2022)]{Bitsch_2022} Bitsch, B., Schneider, A. D., and Kreidberg, L. 2022, A\&A, 665, A138, doi: 10.1051/0004-6361/202243345
\bibitem[Bohn et al.(2020)]{Bohn_2020} Bohn, A. J., Kenworthy, M. A., Ginski, C., et al. 2020, \apjl, 898, L16, doi: 10.3847/2041-8213/aba27e
\bibitem[Boley et al.(2021)]{Boley_2021} Boley, K. M., Wang, J., Zinn, J. C., et al. 2021, AJ, 162, 85, doi: 10.3847/1538-3881/ac0e2d
\bibitem[Boley et al.(2024)]{Boley_2024} Boley, K. M., Christiansen, J. L., Zink, J., et al. 2024, arXiv e-prints, arXiv:2407.13821, doi: 10.48550/arXiv.2407.13821
\bibitem[Bosman et al.(2021)]{Bosman_2021} Bosman, A. D., Alarc´on, F., Bergin, E. A., et al. 2021, \apjs, 257, 7, doi: 10.3847/1538-4365/ac1435
\bibitem[Botelho et al.(2020)]{Botelho_2020} Botelho, R. B., Milone, A. d. C., Mel´endez, J., et al. 2020, \mnras, 499, 2196, doi: 10.1093/mnras/staa2917
\bibitem[Brewer and Fischer(2016)]{Brewer_Fischer_2016} Brewer, J. M., and Fischer, D. A. 2016, \apj, 831, 20, doi: 10.3847/0004-637X/831/1/20
\bibitem[Brewer et al.(2016)]{Brewer_2016} Brewer, J. M., Fischer, D. A., Valenti, J. A., and Piskunov, N. 2016, \apjs, 225, 32, doi: 10.3847/0067-0049/225/2/32
\bibitem[Brooke et al.(2014)]{Brooke_2014} Brooke, J. S. A., Ram, R. S., Western, C. M., et al. 2014, \apjs, 210, 23, doi: 10.1088/0067-0049/210/2/23
\bibitem[Castelli and Kurucz(2003)]{Castelli_Kurucz_2003} Castelli, F., and Kurucz, R. L. 2003, Modelling of Stellar Atmospheres, ed. N. Piskunov, W. W. Weiss, and D. F. Gray, Vol. 210, A20, doi: 10.48550/arXiv.astro-ph/0405087
\bibitem[Coria et al.(2023)]{Coria_2023} Coria, D. R., Crossfield, I. J. M., Lothringer, J., et al. 2023, \apj, 954, 121, doi: 10.3847/1538-4357/acea5f
\bibitem[Costes et al.(2024)]{Costes_2024} Costes, J. C., Xuan, J. W., Vigan, A., et al. 2024, https://arxiv.org/abs/2404.11523
\bibitem[Crossfield(2023)]{Crossfield_2023} Crossfield, I. J. M. 2023, \apjl, 952, L18, doi: 10.3847/2041-8213/ace35f
\bibitem[Crossfield et al.(2019)]{Crossfield_2019} Crossfield, I. J. M., Lothringer, J. D., Flores, B., et al. 2019, \apjl, 871, L3, doi: 10.3847/2041-8213/aaf9b6
\bibitem[de Regt et al.(2024)]{de_Regt_2024} de Regt, S., Gandhi, S., Snellen, I. A. G., et al. 2024, arXiv e-prints, arXiv:2405.10841, doi: 10.48550/arXiv.2405.10841
\bibitem[Delgado Mena et al.(2021)]{Delgado_Mena_CO_2021} Delgado Mena, E., Adibekyan, V., Santos, N. C., et al. 2021, A\&A, 655, A99, doi: 10.1051/0004-6361/202141588
\bibitem[Delrez et al.(2016)]{Delrez_2016} Delrez, L., Santerne, A., Almenara, J. M., et al. 2016, \mnras, 458, 4025, doi: 10.1093/mnras/stw522
\bibitem[Dulick et al.(2003)]{Dulick_2003} Dulick, M., Bauschlicher, C. W., J., Burrows, A., et al. 2003, \apj, 594, 651, doi: 10.1086/376791
\bibitem[Evans et al.(2016)]{Evans_2016} Evans, D. F., Southworth, J., Maxted, P. F. L., et al. 2016, A\&A, 589, A58, doi: 10.1051/0004-6361/201527970
\bibitem[Evans et al.(2018)]{Evans_2018} Evans, T. M., Sing, D. K., Goyal, J. M., et al. 2018, AJ, 156, 283, doi: 10.3847/1538-3881/aaebff
\bibitem[Fabbian et al.(2012)]{Fabbian_2012} Fabbian, D., Moreno-Insertis, F., Khomenko, E., and Nordlund, A. 2012, A\&A, 548, A35, doi: 10.1051/0004-6361/201219335
\bibitem[Finnerty et al.(2024)]{Finnerty_2024} Finnerty, L., Xuan, J. W., Xin, Y., et al. 2024, AJ, 167, 43, doi: 10.3847/1538-3881/ad1180
\bibitem[Fortney(2012)]{Fortney_2012} Fortney, J. J. 2012, \apjl, 747, L27, doi: 10.1088/2041-8205/747/2/L27
\bibitem[Gandhi et al.(2023)]{Gandhi_2023} Gandhi, S., de Regt, S., Snellen, I., et al. 2023, \apjl, 957, L36, doi: 10.3847/2041-8213/ad07e2
\bibitem[Goldman(1982)]{Goldman_1982} Goldman, A. 1982, Appl. Opt., 21, 2100, doi: 10.1364/AO.21.002100
\bibitem[Goorvitch(1994)]{Goorvitch_1994} Goorvitch, D. 1994, \apjs, 95, 535, doi: 10.1086/192110
\bibitem[Gordon et al.(2017)]{Gordon_2017} Gordon, I. E., Rothman, L. S., Hill, C., et al. 2017, Journal of Quantitative Spectroscopy and Radiative Transfer, 203, 3, doi: https://doi.org/10.1016/j.jqsrt.2017.06.038
\bibitem[Grevesse et al.(2007)]{Grevesse_2007} Grevesse, N., Asplund, M., and Sauval, A. J. 2007, SSRv, 130, 105, doi: 10.1007/s11214-007-9173-7
\bibitem[Gustafsson et al.(2008)]{Gustafsson_2008} Gustafsson, Edvardsson, B., Eriksson, K., et al. 2008, A\&A, 486, 951, doi: 10.1051/0004-6361:200809724
\bibitem[Hands and Helled.(2022)]{Hands_Helled_2022} Hands, T. O., and Helled, R. 2022, \mnras, 509, 894, doi: 10.1093/mnras/stab2967
\bibitem[Hawkins et al.(2020)]{Hawkins_2020} Hawkins, K., Lucey, M., Ting, Y.-S., et al. 2020, \mnras, 492, 1164, doi: 10.1093/mnras/stz3132
\bibitem[Hejazi et al.(2023)]{Hejazi_2023} Hejazi, N., Crossfield, I. J. M., Nordlander, T., et al. 2023, \apj, 949, 79, doi: 10.3847/1538-4357/accb97
\bibitem[Hejazi et al.(2024)]{Hejazi_2024} Hejazi, N., Crossfield, I. J. M., Souto, D., et al. 2024, arXiv e-prints, arXiv:2407.07869, doi: 10.48550/arXiv.2407.07869
\bibitem[Hood et al.(2024)]{Hood_2024} Hood, C. E., Mukherjee, S., Fortney, J. J., et al. 2024, arXiv e-prints, arXiv:2402.05345, doi: 10.48550/arXiv.2402.05345
\bibitem[Ilee et al.(2017)]{Ilee_2017} Ilee, J. D., Forgan, D. H., Evans, M. G., et al. 2017, \mnras, 472, 189, doi: 10.1093/mnras/stx1966
\bibitem[Johnson(2010)]{Johnson_2010} Johnson, E. M. 2010, PhD thesis, University of Oregon
\bibitem[Kempton et al.(2018)]{Kempton_2018} Kempton, E. M.-R., Bean, J. L., Louie, D. R., et al. 2018, ASP, 130, 114401, doi: 10.1088/1538-3873/aadf6f
\bibitem[Kobayashi et al.(2020)]{Kobayashi_2020} Kobayashi, C., Karakas, A. I., and Lugaro, M. 2020, arXiv, doi: 10.48550/ARXIV.2008.04660
\bibitem[Kolecki et al.(2022)]{Kolecki_2022} Kolecki, J. R., and Wang, J. 2022, The Astronomical Journal, 164, 87, doi: 10.3847/1538-3881/ac7de3
\bibitem[Kovacs et al.(2019)]{Kovacs_2019} Kovacs, G., and Kovacs, T. 2019, A\&A, 625, A80, doi: 10.1051/0004-6361/201834325
\bibitem[Kurucz et al.(1995)]{Kurucz_1995} Kurucz, R. L. 1995, ASP, Vol. 78, Astrophysical Applications of Powerful New Databases, ed. S. J. Adelman and W. L. Wiese, 205
\bibitem[Lew et al.(2024)]{Lew_2024} Lew, B. W. P., Roellig, T., Batalha, N. E., et al. 2024, arXiv e-prints, arXiv:2402.05900, doi: 10.48550/arXiv.2402.05900
\bibitem[Lincowski et al.(2019)]{Lincowski_2019} Lincowski, A. P., Lustig-Yaeger, J., and Meadows, V. S. 2019, AJ, 158, 26, doi: 10.3847/1538-3881/ab2385
\bibitem[Lincowski et al.(2018)]{Lincowski_2018} Lincowski, A. P., Meadows, V. S., Crisp, D., et al. 2018, \apj, 867, 76, doi: 10.3847/1538-4357/aae36a
\bibitem[Line et al.(2021)]{Line_2021} Line, M. R., Brogi, M., Bean, J. L., et al. 2021, Nature, 598, 580, doi: 10.1038/s41586-021-03912-6
\bibitem[Lopez-Valdivia et al.(2021)]{Lopez_Valdivia_2021} Lopez-Valdivia, R., Sokal, K. R., Mace, G. N., et al. 2021, \apj, 921, 53, doi: 10.3847/1538-4357/ac1a7b
\bibitem[Lothringer et al.(2021)]{Lothringer_2021} Lothringer, J. D., Rustamkulov, Z., Sing, D. K., et al. 2021, \apj, 914, 12, doi: 10.3847/1538-4357/abf8a9
\bibitem[Mace et al.(2018)]{Mace_2018} Mace, G., Sokal, K., Lee, J.-J., et al. 2018, SPIE, Vol. 10702, Ground-based and Airborne Instrumentation for Astronomy VII, ed. C. J. Evans, L. Simard, and H. Takami, 107020Q, doi: 10.1117/12.2312345
\bibitem[Madhusudhan(2012)]{Madhusudhan_2012} Madhusudhan, N. 2012, The Astrophysical Journal, 758, 36, doi: 10.1088/0004-637x/758/1/36
\bibitem[Madhusudhan(2019)]{Madhusudhan_2019} Madhusudhan, N. 2019, A\&A, 57, 617, doi: 10.1146/annurev-astro-081817-051846
\bibitem[Madhusudhan et al.(2014)]{Madhusudhan_2014} Madhusudhan, N., Amin, M. A., and Kennedy, G. M. 2014, \apjl, 794, L12, doi: 10.1088/2041-8205/794/1/L12
\bibitem[Madhusudhan et al.(2017)]{Madhusudhan_2017} Madhusudhan, N., Bitsch, B., Johansen, A., and Eriksson, L. 2017, \mnras, 469, 4102, doi: 10.1093/mnras/stx1139
\bibitem[Mansfield et al.(2022)]{Mansfield_2022} Mansfield, M., Wiser, L., Stevenson, K. B., et al. 2022, AJ, 163, 261, doi: 10.3847/1538-3881/ac658f
\bibitem[Masseron et al.(2014)]{Masseron_2014} Masseron, T., Plez, B., Van Eck, S., et al. 2014, A\&A, 571, A47, doi: 10.1051/0004-6361/201423956
\bibitem[Maxted et al.(2013)]{Maxted_2013} Maxted, P. F. L., Anderson, D. R., Collier Cameron, A., et al. 2013, PASP, 125, 48, doi: 10.1086/669231
\bibitem[Mikal-Evans et al.(2019)]{Mikal-Evans_2019} Mikal-Evans, T., Sing, D. K., Goyal, J. M., et al. 2019, \mnras, 488, 2222, doi: 10.1093/mnras/stz1753
\bibitem[Mikal-Evans et al.(2018)]{Mikal-Evans_2018} Mikal-Evans, T., Sing, D. K., Goyal, J. M., et al. 2018, \aj, 156, 6, 283, doi: 10.3847/1538-3881/aaebff
\bibitem[Molliere et al.(2022)]{Molliere_2022} Molliere, P., Molyarova, T., Bitsch, B., et al. 2022, \apj, 934, 74, doi: 10.3847/1538-4357/ac6a56
\bibitem[Molliere et al.(2019)]{Molliere_2019} Molliere, P., and Snellen, I. A. G. 2019, A\&A, 622, A139, doi: 10.1051/0004-6361/201834169
\bibitem[Mordasini et al.(2016)]{Mordasini_2016} Mordasini, C., van Boekel, R., Molli`ere, P., Henning, T., and Benneke, B. 2016, \apj, 832, 41, doi: 10.3847/0004-637X/832/1/41
\bibitem[Mortier et al.(2013)]{Mortier_2013} Mortier, A., Santos, N. C., Sousa, S., et al. 2013, A\&A, 551, A112, doi: 10.1051/0004-6361/201220707
\bibitem[Morton et al.(2015)]{Morton_2015} Morton, T. D. 2015, isochrones: Stellar model grid package, Astrophysics Source Code Library, record ascl:1503.010
\bibitem[Nissen et al.(2018)]{Nissen_Gustafsson_2018} Nissen, P. E., and Gustafsson, B. 2018, A\&A Rv, 26, 6, doi: 10.1007/s00159-018-0111-3
\bibitem[Nomura et al.(2023)]{Nomura_2023} Nomura, H., Furuya, K., Cordiner, M. A., et al. 2023, ASP, Vol. 534, Protostars and Planets VII, ed. S. Inutsuka, Y. Aikawa, T. Muto, K. Tomida, and M. Tamura, 1075
\bibitem[Oberg et al.(2011)]{Oberg_2011} Oberg, K. I., Murray-Clay, R., and Bergin, E. A. 2011, AJ, 743, L16, doi: 10.1088/2041-8205/743/1/l16
\bibitem[Pacetti et al.(2022)]{Pacetti_2022} Pacetti, E., Turrini, D., Schisano, E., et al. 2022, \apj, 937, 36, doi: 10.3847/1538-4357/ac8b11
\bibitem[Park et al.(2014)]{Park_2014} Park, C., Jaffe, D. T., Yuk, I.-S., et al. 2014, SPIE, Vol. 9147, Ground-based and Airborne Instrumentation for Astronomy V, ed. S. K. Ramsay, I. S. McLean, and H. Takami, 91471D, doi: 10.1117/12.2056431
\bibitem[Petigura et al.(2018)]{Petigura_2018} Petigura, E. A., Marcy, G. W., Winn, J. N., et al. 2018, AJ, 155, 89, doi: 10.3847/1538-3881/aaa54c
\bibitem[Petrovich et al.(2015a)]{Petrovich_2015a} Petrovich, C. 2015a, \apj, 805, 75, doi: 10.1088/0004-637X/805/1/75
\bibitem[Petrovich et al.(2015b)]{Petrovich_2015b} Petrovich, C. 2015b, \apj, 799, 27, doi: 10.1088/0004-637X/799/1/27
\bibitem[Plez et al.(2012)]{Plez_2012} Plez, B. 2012, Turbospectrum: Code for spectral synthesis, Astrophysics Source Code Library, record ascl:1205.004
\bibitem[Polanski et al.(2022)]{Polanski_2022} Polanski, A. S., Crossfield, I. J. M., Howard, A. W., Isaacson, H., and Rice, M. 2022, https://arxiv.org/abs/2207.13662
\bibitem[Pollacco et al.(2006)]{Pollacco_2006} Pollacco, D. L., Skillen, I., Collier Cameron, A., et al. 2006, PASP, 118, 1407, doi: 10.1086/508556
\bibitem[Prantzos et al.(2018)]{Prantzos_2018} Prantzos, N., Abia, C., Limongi, M., Chieffi, A., and Cristallo, S. 2018, \mnras, 476, 3432, doi: 10.1093/mnras/sty316
\bibitem[Reggiani et al.(2024)]{Reggiani_2024} Reggiani, H., Galarza, J. Y., Schlaufman, K. C., et al. 2024, AJ, 167, 45, doi: 10.3847/1538-3881/ad0f93
\bibitem[Reggiani et al.(2022)]{Reggiani_2022} Reggiani, H., Schlaufman, K. C., Healy, B. F., Lothringer, J. D., and Sing, D. K. 2022, AJ, 163, 159, doi: 10.3847/1538-3881/ac4d9f
\bibitem[Romano(2022)]{Romano_2022} Romano, D. 2022, The Astronomy and Astrophysics Review, 30, doi: 10.1007/s00159-022-00144-z
\bibitem[Romano et al.(2020)]{Romano_2020} Romano, D., Franchini, M., Grisoni, V., et al. 2020, A\&A, 639, A37, doi: 10.1051/0004-6361/202037972
\bibitem[Rosenthal et al.(2021)]{Rosenthal_2021} Rosenthal, L. J., Fulton, B. J., Hirsch, L. A., et al. 2021, \apjs, 255, 8, doi: 10.3847/1538-4365/abe23c
\bibitem[Rothman et al.(2021)]{Rothman_2021} Rothman, L. S. 2021, Nature Reviews Physics, 3, 302, doi: 10.1038/s42254-021-00309-2
\bibitem[Ryabchikova et al.(2015)]{Ryabchikova_2015} Ryabchikova, T., Piskunov, N., Kurucz, R. L., et al. 2015, PhyS, 90, 054005, doi: 10.1088/0031-8949/90/5/054005
\bibitem[Ryabchikova et al.(2022)]{Ryabchikova_2022} Ryabchikova, T., Piskunov, N., and Pakhomov, Y. 2022, Atoms, 10, doi: 10.3390/atoms10040103
\bibitem[Schneider and Bitsch(2021a)]{Schneider_Bitsch_2021a} Schneider, A. D., and Bitsch, B. 2021a, A\&A, 654, A71, doi: 10.1051/0004-6361/202039640
\bibitem[Schneider and Bitsch(2021b)]{Schneider_Bitsch_2021b} Schneider, A. D., and Bitsch, B.. 2021b, A\&A, 654, A72, doi: 10.1051/0004-6361/202141096
\bibitem[Seligman et al.(2022)]{Seligman_2022} Seligman, D. Z., Rogers, L. A., Cabot, S. H. C., et al. 2022, PSJ, 3, 150, doi: 10.3847/PSJ/ac75b5
\bibitem[Shchukina et al.(2016)]{Shchukina_2016} Shchukina, N., Sukhorukov, A., and Trujillo Bueno, J. 2016, A\&A, 586, A145, doi: 10.1051/0004-6361/201526452
\bibitem[Sing et al.(2019)]{Sing_2019} Sing, D. K., Lavvas, P., Ballester, G. E., et al. 2019, AJ, 158, 91, doi: 10.3847/1538-3881/ab2986
\bibitem[Smith et al.(2015)]{Smith_2015} Smith, R. L., Pontoppidan, K. M., Young, E. D., and Morris, M. R. 2015, \apj, 813, 120, doi: 10.1088/0004-637x/813/2/120
\bibitem[Sneden et al.(2012)]{Sneden_2012} Sneden, C., Bean, J., Ivans, I., Lucatello, S., and Sobeck, J. 2012, MOOG: LTE line analysis and spectrum synthesis, Astrophysics Source Code Library, record ascl:1202.009
\bibitem[Sneden et al.(2014)]{Sneden_2014} Sneden, C., Lucatello, S., Ram, R. S., Brooke, J. S. A., and Bernath, P. 2014, \apjs, 214, 26, doi: 10.1088/0067-0049/214/2/26
\bibitem[Sneden(1973)]{Sneden_1973} Sneden, C. A. 1973, PhD thesis, University of Texas, Austin
\bibitem[Souto et al.(2017)]{Souto_2017} Souto, D., Cunha, K., Garc´ıa-Hern´andez, D. A., et al. 2017, \apj, 835, 239, doi: 10.3847/1538-4357/835/2/239
\bibitem[Souto et al.(2018)]{Souto_2018} Souto, D., Unterborn, C. T., Smith, V. V., et al. 2018, \apjl, 860, L15, doi: 10.3847/2041-8213/aac896
\bibitem[Souto et al.(2022)]{Souto_2022} Souto, D., Cunha, K., Smith, V. V., et al. 2022, \apj, 927, 123, doi: 10.3847/1538-4357/ac4891
\bibitem[Ting et al.(2018)]{Ting_2018} Ting, Y.-S., Conroy, C., Rix, H.-W., and Asplund, M. 2018, \apj, 860, 159, doi: 10.3847/1538-4357/aac6c9
\bibitem[Turrini et al.(2021)]{Turrini_2021} Turrini, D., Schisano, E., Fonte, S., et al. 2021, \apj, 909, 40, doi: 10.3847/1538-4357/abd6e5
\bibitem[Unterborn et al.(2017)]{Unterborn_2017} Unterborn, C. T., Hull, S. D., Stixrude, L. P., et al. 2017, arXiv, https://arxiv.org/abs/1706.10282
\bibitem[Unterborn et al.(2014)]{Unterborn_2014} Unterborn, C. T., Kabbes, J. E., Pigott, J. S., Reaman, D. M., and Panero, W. R. 2014, \apj, 793, 124, doi: 10.1088/0004-637x/793/2/124
\bibitem[Valenti and Fischer(2005)]{Valenti_Fischer_2005} Valenti, J. A., and Fischer, D. A. 2005, in Protostars and Planets V Posters, 8592
\bibitem[Vogt et al.(1994)]{Vogt_1994} Vogt, S. S., Allen, S. L., Bigelow, B. C., et al. 1994, SPIE, Vol. 2198, Instrumentation in Astronomy VIII, ed. D. L. Crawford and E. R. Craine, 362, doi: 10.1117/12.176725
\bibitem[Wang and Fischer(2015)]{Wang_Fischer_2015} Wang, J., and Fischer, D. A. 2015, AJ, 149, 14, doi: 10.1088/0004-6256/149/1/14
\bibitem[Woods(2009)]{Woods_2009} Woods, P. M. 2009, arXiv e-prints, arXiv:0901.4513, doi: 10.48550/arXiv.0901.4513
\bibitem[Woods and Willacy(2009)]{Woods_Willacy_2009} Woods, P. M., and Willacy, K. 2009, \apj, 693, 1360, doi: 10.1088/0004-637X/693/2/1360
\bibitem[Wu and Lithwick(2011)]{Wu_Lithwick_2011} Wu, Y., and Lithwick, Y. 2011, \apj, 735, 109, doi: 10.1088/0004-637X/735/2/109
\bibitem[Wu and Murray(2003)]{Wu_Murray_2003} Wu, Y., and Murray, N. 2003, \apj, 589, 605, doi: 10.1086/374598
\bibitem[Xuan et al.(2024a)]{Xuan_2024} Xuan, J. W., Wang, J., Finnerty, L., et al. 2024, \apj, 962, 10, doi: 10.3847/1538-4357/ad1243
\bibitem[Xuan et al.(2024b)]{Xuan_2024b} Xuan, J. W., Hsu, C.-C., Finnerty, L., et al. 2024, arXiv e-prints, arXiv:2405.13128, doi: 10.48550/arXiv.2405.13128
\bibitem[Yoshida et al.(2024)]{Yoshida_2024} Yoshida, T. C., Nomura, H., Furuya, K., et al. 2024, arXiv e-prints, arXiv:2403.00626, doi: 10.48550/arXiv.2403.00626
\bibitem[Zhang et al.(2017)]{Zhang_2017} Zhang, K., Bergin, E. A., Blake, G. A., Cleeves, L. I., and Schwarz, K. R. 2017, Nature Astronomy, 1, 0130, doi: 10.1038/s41550-017-0130
\bibitem[Zhang et al.(2021a)]{Zhang_2021a} Zhang, Y., Snellen, I. A. G., and Molliere, P. 2021a, Astronomy and Astrophysics, 656, A76, doi: 10.1051/0004-6361/202141502
\bibitem[Zhang et al.(2021b)]{Zhang_2021b} Zhang, Y., Snellen, I. A. G., Bohn, A. J., et al. 2021b, Nature, 595, 370, doi: 10.1038/s41586-021-03616-x
\bibitem[Zink and Howard(2023)]{Zink_2023} Zink, J. K., and Howard, A. W. 2023, \apjl, 956, L29, doi: 10.3847/2041-8213/acfdab

\end{thebibliography}



\end{document}